\documentclass[11pt,a4paper]{article}

\usepackage[margin=1in]{geometry}
\usepackage{graphicx}
\usepackage{booktabs}
\usepackage{amsmath}
\usepackage{siunitx}
\usepackage{xcolor}
\usepackage{authblk}
\usepackage{longtable}
\usepackage{float}
\usepackage{placeins}
\usepackage{hyperref}
\usepackage{cleveref}
\newcommand{\coreDeltaFix}{\ensuremath{-0.92}}

\newcommand{\coreRecovery}{0.989}

\newcommand{\coreRecoveryRatioOfMeans}{0.962}

\newcommand{\epochCellsWord}{eight}
\newcommand{\epochRecords}{28}
\newcommand{\epochEvaluations}{580}
\newcommand{\epochAachenEvaluations}{240}
\newcommand{\epochPittsburghEvaluations}{340}
\newcommand{\allFixedEvaluations}{1460}
\newcommand{\allResilienceZeroEvaluations}{1256}

\newcommand{\epochMaxCellDselChange}{2.21}
\newcommand{\epochMaxBestShift}{29.08}
\newcommand{\epochResolutionRange}{0.60--9.06}

\newcommand{\fixedQPUSeconds}{4256.0}
\newcommand{\studyQPUSeconds}{8871.0}

\newcommand{\epochMeanAbsCellDselChange}{1.34}

\newcommand{\epochMaxFinalShift}{31.29}

\newcommand{\epochMaxChangeOverResolution}{1.04}
\newcommand{\epochPrimaryDrift}{23.99}
\newcommand{\epochMaxChangePctDrift}{9.2}
\newcommand{\epochLagMin}{1.5}
\newcommand{\epochLagMax}{9.0}
\newcommand{\aerPairCoreRecords}{29}

\newcommand{\aerPairMaxCoreGap}{0.116}

\newcommand{\clusterCoreRecoveryLow}{0.906}
\newcommand{\clusterCoreRecoveryHigh}{1.072}

\newcommand{\clusterCoreRatioLow}{0.853}
\newcommand{\clusterCoreRatioHigh}{1.175}

\newcommand{\clusterExtendedRatioLow}{1.005}
\newcommand{\clusterExtendedRatioHigh}{1.218}

\newcommand{\idealGapMedian}{0.0118}
\newcommand{\idealGapMin}{0.00042}
\newcommand{\idealGapMax}{1.415}

\newcommand{\subCasciCountWord}{Seven}

\newcommand{\subCasciMaximum}{17.36}

\newcommand{\fleetLayerSpan}{3.24}
\newcommand{\classLayerSpan}{1.65}
\newcommand{\fleetReadoutSpan}{5.80}
\newcommand{\classReadoutSpan}{3.72}
\newcommand{\phoenixClopsRatio}{6.06}
\newcommand{\phoenixMcpsRatio}{28.87}
\newcommand{\phoenixReadoutRatio}{2.20}
\newcommand{\pittsburghLayerRank}{2/6}
\newcommand{\aachenLayerRank}{4/6}
\newcommand{\sigmaMin}{1.72}
\newcommand{\sigmaMax}{58.18}

\newcommand{\rOneClopsLow}{13.2}
\newcommand{\rOneClopsHigh}{13.8}
\newcommand{\rOneMcps}{15.2}
\newcommand{\rOneLayerLow}{1.3}
\newcommand{\rOneLayerHigh}{3.1}
\newcommand{\rOneReadoutLow}{3.6}
\newcommand{\rOneReadoutHigh}{5.5}

\newcommand{\phoenixDselOverB}{2.04}
\newcommand{\phoenixFixedQPU}{12.0}
\newcommand{\phoenixFixedRate}{0.12}
\newcommand{\phoenixSourceQPU}{315.0}
\newcommand{\phoenixSourceRate}{3.00}
\newcommand{\phoenixDuplicateQPU}{12.0}
\newcommand{\phoenixAllQPU}{339.0}
\newcommand{\phoenixRatioOfMeans}{2.11}

\newcommand{\phoenixPlateauDselRatio}{1.25}
\newcommand{\phoenixPlateauDselRatioMeans}{1.17}

\newcommand{\coreRecoveryLooLow}{0.953}
\newcommand{\coreRecoveryLooHigh}{1.021}
\newcommand{\coreRecoveryPrecisionWeighted}{0.892}

\newcommand{\exploratoryPermutations}{50,000}
\newcommand{\competitorRunGainMin}{31.9}
\newcommand{\competitorRunGainMax}{56.4}
\newcommand{\competitorCellGainMin}{31.9}
\newcommand{\competitorCellGainMax}{49.4}

\newcommand{\precisionHtwoCount}{182}
\newcommand{\precisionWaterCount}{3964}

\newcommand{\inclusionBase}{0.989}
\newcommand{\inclusionPermissive}{1.007}
\newcommand{\inclusionStrict}{0.990}

\newcommand{\plateauRatioMean}{1.025}
\newcommand{\plateauRatioSD}{0.527}
\newcommand{\plateauRatioCILow}{0.690}
\newcommand{\plateauRatioCIHigh}{1.360}

\newcommand{\debiasedUnderCount}{6}
\newcommand{\debiasedOverCount}{2}

\newcommand{\corrLiH}{0.27}
\newcommand{\corrHF}{0.76}
\newcommand{\corrOtwo}{0.75}

\newcommand{\tieLatticeMatches}{306}
\newcommand{\tieLatticeEvaluations}{306}
\newcommand{\tieLatticePairOpportunities}{3217}
\newcommand{\tieConditionalProbability}{0.0728}

\graphicspath{{figures/}}
\hypersetup{
    colorlinks=true,
    linkcolor=blue!55!black,
    citecolor=blue!55!black,
    urlcolor=blue!55!black
}

\title{The winner's curse in hardware VQE: drift-differenced remeasurement of finite-shot selection bias}
\author[1]{Julen Larrucea\thanks{Correspondence: \href{mailto:julen.larrucea@ifam.fraunhofer.de}{julen.larrucea@ifam.fraunhofer.de}; ORCID: \href{https://orcid.org/0000-0003-0582-9259}{0000-0003-0582-9259}}}
\affil[1]{Fraunhofer Institute for Manufacturing Technology and Advanced Materials IFAM, Wiener Strasse 12, 28359 Bremen, Germany}
\date{}

\begin{document}
\maketitle

\begin{abstract}
Finite-shot optimization can make noisy variational quantum eigensolver energies look artificially accurate. We studied six hardware-tractable molecular active spaces on two IBM Heron r3 processors. Seven optimizer-selected values fell below the exact active-space eigenvalue by up to \subCasciMaximum{} mHa. The cell-balanced fraction of the source-observed best--final advantage not retained on later fixed-parameter measurement was \coreRecovery{} (approximate 95\% session-level interval $[\clusterCoreRecoveryLow{},\clusterCoreRecoveryHigh{}]$); alternative aggregations span \coreRecoveryPrecisionWeighted{}--\coreRecovery{}. Thus little of the selected advantage remained on average. Measurements occurred 5.6--38.4 days later, so parameter-dependent drift remains a competing explanation. Best-observed hardware VQE values require prompt independent confirmation before supporting accuracy claims.
\end{abstract}

\noindent\textbf{Subject terms/keywords:} variational quantum eigensolver; quantum chemistry; hardware benchmarking; reproducibility; error mitigation; superconducting qubits.

\section*{Introduction}

Variational quantum eigensolvers combine a parameterized quantum circuit with a classical optimizer to approximate molecular ground-state energies \cite{peruzzo2014vqe,kandala2017hardware,tilly2022vqe,bharti2022nisq}.
For current noisy devices, the decisive experimental question is not only whether a molecule can be encoded, but whether the resulting circuit is shallow, stable, reproducible, and sufficiently documented to support quantitative conclusions.
Active-space reduction and qubit tapering are central ingredients: they lower the qubit and gate requirements, but also define the physical model being benchmarked.
Recent VQE work calls selection of the lowest noisy trajectory value a winner's curse \cite{novak2025reliable}. We examine that risk on two Heron processors by comparing selected and final parameter points with later fixed-parameter measurements. Differencing the two shifts removes a change common to both points, but cannot remove parameter-dependent drift.

Early superconducting H$_2$ VQE work \cite{omalley2016scalable} led to broader chemistry benchmarks and measurement-resource studies \cite{mccaskey2019benchmark,arute2020hartree,gonthier2022measurements}. The present study is not another chemical-accuracy benchmark: its molecule ladder makes the reporting problem visible across circuit sizes. It retains complete trajectories, later measurements of both saved parameter points, circuit resources and calibration summaries. This follows the principle that a hardware benchmark should expose the computational stack and avoid misleading performance metrics \cite{proctor2024benchmarking,proctor2020capabilities}.

Cross-platform suites, error-mitigation methods and alternative eigensolvers address different aspects of hardware performance \cite{lubinski2023qedc,tomesh2022supermarq,temme2017error,vandenberg2022trex,robledomoreno2024chemistry}. Here the narrower question is whether the apparent accuracy of an optimizer-selected hardware energy survives independent measurement.

\textbf{Relation to the prior H$_2$ study.} An earlier study by the author and co-workers \cite{julen2026h2bench} focused on H$_2$. The present work extends that protocol to six molecules and two Heron backends and adds repeated runs together with fixed-parameter remeasurement.
A source-constrained scope comparison with the closest prior studies is provided in Supplementary Table~\ref{tab:prior-work}.

The primary endpoint is the fraction of the source-observed best--final advantage not retained at the later measurement epoch. We estimate it across eight replicated molecule/backend cells and compare the scale with an empirical order-statistic calculation. Four single-source deep-circuit cells are descriptive extensions, not additional replication.
The endpoint quantifies loss of an observed advantage across epochs; it does not identify how much of that loss was caused by selection rather than by a parameter-dependent change between acquisition epochs.

\section*{Results}
\subsection*{Workloads and circuit resources}

\begin{table}[H]
\centering
\scriptsize
\resizebox{\linewidth}{!}{%
\begin{tabular}{llrrrrrrr}
\toprule
Molecule & Backend & $n$ & Mean best $E$ (Ha) & Mean final $E$ (Ha) & $\sigma_\mathrm{best}$ (mHa) & SEM (mHa) & 95\% CI half-width (mHa) & Signed best error (mHa) \\
\midrule
\multicolumn{9}{l}{\textit{Replicated suite}} \\
H$_2$ & Aachen & 3 & -1.85302 & -1.83771 & 5.46 & 3.15 & $\pm$ 13.56 & -3.36 \\
 & Pittsburgh & 4 & -1.85394 & -1.83587 & 17.12 & 8.56 & $\pm$ 27.24 & -4.28 \\
LiH & Aachen & 4 & -1.04981 & -1.04610 & 0.62 & 0.31 & $\pm$ 0.99 & +3.42 \\
 & Pittsburgh & 4 & -1.03656 & -1.03283 & 1.72 & 0.86 & $\pm$ 2.74 & +16.67 \\
HF & Aachen & 3 & -1.92576 & -1.92057 & 2.29 & 1.32 & $\pm$ 5.69 & -0.175 \\
 & Pittsburgh & 4 & -1.92060 & -1.90879 & 5.51 & 2.75 & $\pm$ 8.76 & +4.98 \\
O$_2$ & Aachen & 2 & -4.89627 & -4.87415 & 16.71 & 11.81 & $\pm$ 150.09 & +466.38 \\
 & Pittsburgh & 5 & -4.97033 & -4.86436 & 77.96 & 34.87 & $\pm$ 96.80 & +392.32 \\
\addlinespace
\multicolumn{9}{l}{\textit{Single-record stress tests}} \\
BeH$_2$ & Aachen & 1 & -2.94234 & -2.86338 & n/a & n/a & n/a & +975.41 \\
 & Pittsburgh & 1 & -2.81439 & -2.74552 & n/a & n/a & n/a & +1103.36 \\
H$_2$O & Aachen & 1 & -4.19782 & -4.00723 & n/a & n/a & n/a & +1948.00 \\
 & Pittsburgh & 1 & -4.20137 & -4.14702 & n/a & n/a & n/a & +1944.44 \\
\bottomrule
\end{tabular}}
\caption{Hardware active-space energies are in hartree; dispersion, interval half-width and signed error relative to active-space CASCI are in mHa. Best-observed and final means are distinct. SEM is $\sigma/\sqrt{n}$ and the interval uses a two-sided $t_{n-1}$ critical value. Stress-test uncertainty is unavailable for $n=1$.}
\label{tab:coverage-replicated}\label{tab:coverage-stress}
\end{table}

\Cref{tab:coverage-replicated} summarizes the successful hardware coverage.
The replicated core dataset consists of H$_2$, LiH, HF, and O$_2$ on both \texttt{ibm\_aachen} and \texttt{ibm\_pittsburgh}.
In tables and figures, Aachen and Pittsburgh denote \texttt{ibm\_aachen} and \texttt{ibm\_pittsburgh}; A/P gives the same order. This subset is the most suitable for backend-to-backend comparison because it contains repeated independent runs.
BeH$_2$ and H$_2$O were measured once on each backend and are retained as larger-circuit stress tests, without replicated statistical comparisons.
The record dispositions are consolidated in Supplementary Table~\ref{tab:inclusion-flow}. Including the zero-source-gap Pittsburgh H$_2$ record in its cell-level ratio gives a permissive non-retained fraction of \inclusionPermissive{}; dropping that entire cell as a strict source-integrity sensitivity gives \inclusionStrict{}, compared with the primary \inclusionBase{}. The zero-gap record has no meaningful record-level ratio, and neither sensitivity removes the source-to-fixed epoch confound.

\begin{table}[H]
\centering
\small
\begin{tabular}{lllllll}
\toprule
Molecule & Active & Logical q. & Params & Depth A/P & 2q gates A/P & Ham. terms \\
\midrule
H$_2$ & 2e2o & 1 & 1 & 5/5 & 0/0 & 3 \\
LiH & 2e2o & 2 & 1 & 43/43 & 4/4 & 9 \\
HF & 2e2o & 1 & 1 & 5/5 & 0/0 & 3 \\
O$_2$ & 4e4o & 4 & 1 & 74/74 & 12/12 & 50 \\
BeH$_2$ & 4e4o & 5 & 8 & 1935/1979 & 391/451 & 119 \\
H$_2$O & 4e4o & 5 & 8 & 2275/2131 & 602/542 & 155 \\
\bottomrule
\end{tabular}
\caption{As-executed circuit resources at optimization level 0 for parity mapping with particle-number tapering (PT) and UCC doubles. A/P denotes Aachen/Pittsburgh.}
\label{tab:resources}
\end{table}

H$_2$ and HF taper to one logical qubit; LiH uses two, and O$_2$ uses four with 12 two-qubit gates (\Cref{tab:resources}). BeH$_2$ and H$_2$O instead require eight parameters and hundreds of two-qubit gates. These are level-0, routed circuits, not optimized deployment costs. Circuit size was chosen for hardware tractability rather than for a uniform chemistry benchmark.

Used-resource calibration summaries are in Supplementary Table~\ref{tab:hardware}; H$_2$ and HF have no two-qubit edge.

\subsection*{Selected minima and state controls}

The backend comparison in \Cref{tab:coverage-replicated,tab:coverage-stress} should be interpreted as a hardware reproducibility benchmark for fixed active-space Hamiltonians and fixed circuit construction, not as a statement of absolute chemical accuracy.
For H$_2$, LiH, and HF, repeated runs on both backends produce compact energy ranges relative to the larger O$_2$ scatter on \texttt{ibm\_pittsburgh}.
O$_2$ remains valuable because it adds an open-shell active-space example with modest circuit depth.
The increased spread of O$_2$ illustrates that low gate count alone does not guarantee optimizer stability or backend-independent estimates.
The single-run Pittsburgh BeH$_2$ and H$_2$O measurements complete stress-test coverage on both Heron processors; despite lower recorded readout errors than the corresponding \texttt{ibm\_aachen} stress-test runs, their errors remain approximately 1.10 Ha and 1.94 Ha relative to the active-space CASCI references.
Only successful runs are summarized; failed attempts and superseded submissions are excluded from quantitative endpoints.
There are 33 molecule-level records from 17 distinct hardware sessions in the primary dataset after the exclusions described in Methods. Their optimization traces are shown in Supplementary Figure~\ref{fig:optimizer-traces}.
For LiH, HF, and O$_2$, active-space correlation magnitudes are \corrLiH{}, \corrHF{}, and \corrOtwo{} mHa, below the 1.594 mHa chemical-accuracy threshold. Their cells mainly test estimator reproducibility. The O$_2$ 4e4o window also breaks bonding $\pi_u$ degeneracy, so its CASCI reference is symmetry broken. BeH$_2$ and H$_2$O remain roughly 1--2 Ha from their active-space references. This is a circuit-size ladder, not a demonstration of chemical accuracy.

\subCasciCountWord{} optimizer-selected run energies are below the exact active-space eigenvalue by up to \subCasciMaximum{} mHa (Supplementary Table~\ref{tab:sub-casci-audit}). They are finite-shot excursions, not physical variational energies. A single noisy evaluation can also fall below the bound, so these observations alone do not measure how much additional bias selecting the minimum causes.

In all \epochRecords{} runs entering the replicated nonzero-source-gap contrast, the saved best and final vectors differ by $\lVert\Delta\theta\rVert_2=1.0\times10^{-4}$ in one variational parameter (Supplementary Table~\ref{tab:ideal-gap-control}). The remaining Pittsburgh H$_2$ core record differs by $1.0\times10^{-3}$, has zero source gap and is excluded from that contrast. Thus selection compares nearly identical parameter settings, not widely separated candidate states. Their proximity alone does not bound a parameter-dependent hardware response or prove that the observed source gap is exclusively min-of-trajectory bias.

Noiseless evaluation at the archived best and final parameters gives a median absolute gap of \idealGapMedian{} mHa across all 33 source runs. Under the compact noisy Aer models, the record-level maximum absolute best--final gap is \aerPairMaxCoreGap{} mHa among the \aerPairCoreRecords{} replicated-cell records, as recorded in \texttt{fixed\_theta\_pair\_aer.json} (Supplementary Table~\ref{tab:fixed-theta-pair-aer}). These model-based controls support near-equivalence of the saved states under the tested Hamiltonians and noise models; they do not exclude a parameter-dependent device effect. The broader simulation comparison is in Supplementary Table~\ref{tab:simulation-baselines} and Figure~\ref{fig:simulation-baselines}.

\begin{table}[!htbp]
\centering
\scriptsize
\resizebox{\linewidth}{!}{%
\begin{tabular}{lrrrrrrr}
\toprule
Molecule & $n$ & Source error & Fixed error & Shift & 95\% CI & Paired $p$ & Holm $p$ \\
\midrule
\multicolumn{8}{l}{\textbf{Aachen}} \\
H$_2$ & 3 & $-3.36$ & +13.77 & +17.13 & $\pm$ 8.45 & 0.0129 & 0.0397 \\
LiH & 4 & +3.42 & +7.79 & +4.37 & $\pm$ 2.38 & 0.0099 & 0.0397 \\
HF & 3 & $-0.175$ & +9.77 & +9.94 & $\pm$ 5.47 & 0.0160 & 0.0397 \\
O$_2$ & 2 & +466.38 & +588.57 & +122.19 & $\pm$ 262.29 & 0.1065 & 0.1065 \\
BeH$_2$ & 1 & +975.41 & +1348.41 & +373.00 & n/a & n/a & n/a \\
H$_2$O & 1 & +1948.00 & +2002.73 & +54.73 & n/a & n/a & n/a \\
\midrule
\multicolumn{8}{l}{\textbf{Pittsburgh}} \\
H$_2$ & 4 & $-4.28$ & +25.63 & +29.92 & $\pm$ 24.00 & 0.0286 & 0.0572 \\
LiH & 4 & +16.67 & +10.46 & $-6.20$ & $\pm$ 3.07 & 0.0076 & 0.0280 \\
HF & 4 & +4.98 & +21.71 & +16.73 & $\pm$ 8.04 & 0.0070 & 0.0280 \\
O$_2$ & 5 & +392.32 & +383.07 & $-9.25$ & $\pm$ 86.91 & 0.7824 & 0.7824 \\
BeH$_2$ & 1 & +1103.36 & +1185.09 & +81.72 & n/a & n/a & n/a \\
H$_2$O & 1 & +1944.44 & +2068.45 & +124.01 & n/a & n/a & n/a \\
\bottomrule
\end{tabular}%
}
\caption{Independent unmitigated remeasurement of optimizer-selected $\theta_\mathrm{best}$ vectors. Values are in mHa relative to active-space CASCI and positive shift denotes a higher remeasured energy. Intervals and paired tests use one mean per source record. Holm correction is applied separately to the four core tests on each backend; single-record stress tests have no run-level inference. Pittsburgh H$_2$ uses $n=4$ here, giving a best-point shift of $+29.92$ mHa; the analysis-defined $d_{\rm sel}$ endpoint excludes its zero-source-gap diagnostic ($+8.411$ mHa best shift and $-1.994$ mHa final shift), leaving $n=3$, $+37.08$ mHa and $+13.42$ mHa, respectively.}
\label{tab:fixed-validation}
\end{table}

\begin{table}[t]
\centering
\scriptsize
\resizebox{\linewidth}{!}{%
\begin{tabular}{lrrrrrrr}
\toprule
Molecule & $n$ & $\mathrm{drift}_{src}$ & $\Delta_{\rm fix}$ & $d_\mathrm{sel}$ & $f_{\rm nr}$ & Empirical $b$ & $d_\mathrm{sel}/b$ \\
\midrule
\multicolumn{8}{l}{\textbf{Aachen}} \\
H$_2$ & 3 & +15.31 & -1.79 & +13.52 & 0.883 & 10.73 & 1.26 \\
LiH & 4 & +3.71 & +0.09 & +3.79 & 1.023 & 3.71 & 1.02 \\
HF & 3 & +5.19 & +1.25 & +6.44 & 1.241 & 7.21 & 0.89 \\
O$_2$ & 2 & +22.12 & +3.00 & +25.13 & 1.136 & 35.00 & 0.72 \\
BeH$_2$ & 1 & +78.96 & -7.80 & +71.16 & 0.901 & 46.75 & 1.52 \\
H$_2$O & 1 & +190.59 & +19.68 & +210.27 & 1.103 & 117.65 & 1.79 \\
\midrule
\multicolumn{8}{l}{\textbf{Pittsburgh}} \\
H$_2$ & 3 & +24.09 & -0.43 & +23.67 & 0.982 & 22.43 & 1.06 \\
LiH & 4 & +3.73 & -0.21 & +3.52 & 0.945 & 4.15 & 0.85 \\
HF & 4 & +11.82 & -2.80 & +9.02 & 0.763 & 11.79 & 0.76 \\
O$_2$ & 5 & +105.97 & -6.47 & +99.50 & 0.939 & 64.29 & 1.55 \\
BeH$_2$ & 1 & +68.87 & -6.70 & +62.17 & 0.903 & 40.69 & 1.53 \\
H$_2$O & 1 & +54.35 & -8.48 & +45.88 & 0.844 & 118.71 & 0.39 \\
\bottomrule
\end{tabular}}
\caption{Decomposition $d_\mathrm{sel}=\mathrm{drift}_{src}+\Delta_{\rm fix}$ and Gaussian order-statistic scale in mHa. Here $\mathrm{drift}_{src}=E_{\rm src}(\theta_{\rm final})-E_{\rm src}(\theta_{\rm best})$, $\Delta_{\rm fix}=E_{\rm fix}(\theta_{\rm best})-E_{\rm fix}(\theta_{\rm final})$, and the non-retained fraction is $f_{\rm nr}=d_\mathrm{sel}/\mathrm{drift}_{src}$. If the source-observed advantage were retained under independent measurement, $\Delta_{\rm fix}$ would equal $-\mathrm{drift}_{src}$; $\Delta_{\rm fix}=0$ instead means that the two fixed energies are equal on average. The empirical scale uses Blom's approximation \cite{blom1958statistical}. The Pittsburgh H$_2$ selection set uses three records; its fourth zero-source-drift record is reported as a zero-source-gap diagnostic. Energy columns and $b$ are means over source records; displayed ratios are ratios of those means, and $n$ counts source records. Stress cells are descriptive. Columns are rounded independently; the identity $d_\mathrm{sel}=\mathrm{drift}_{src}+\Delta_{\rm fix}$ closes exactly on unrounded values.}
\label{tab:dsel-primary}
\end{table}

\begin{figure}[H]
    \centering
    \includegraphics[width=\linewidth]{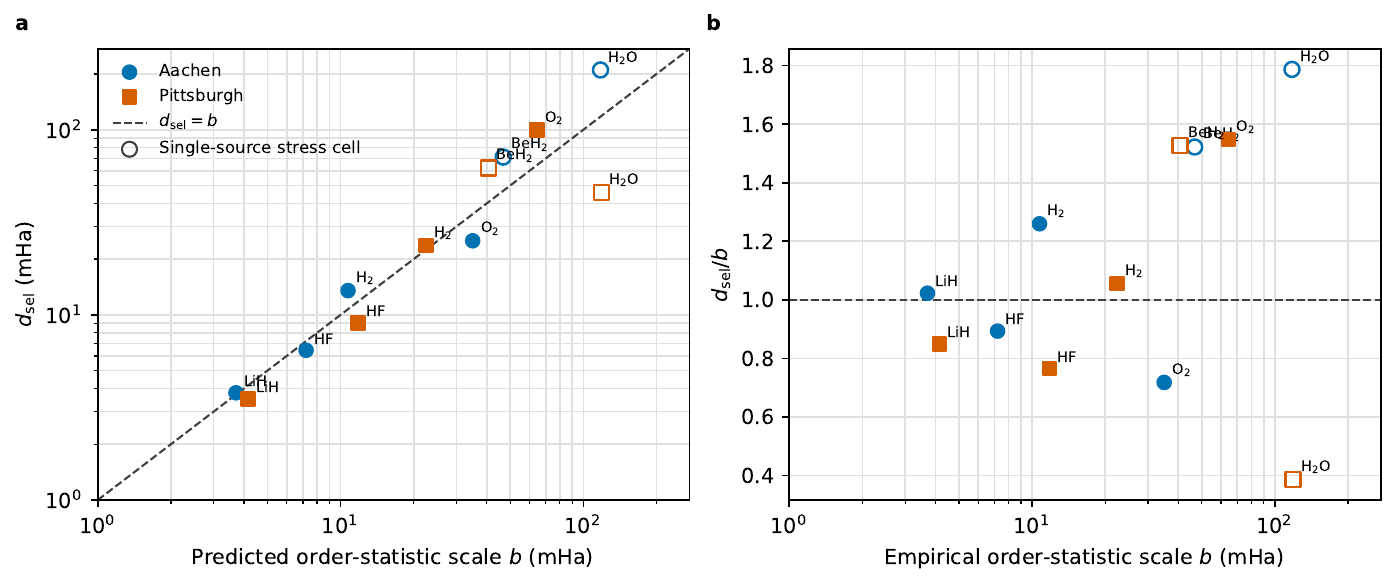}
    \caption{Drift-differenced contrast and its scale-normalized form. \textbf{a}, $d_{\rm sel}$ against the empirical Gaussian order-statistic scale $b=a_m s_b$, where $a_m$ follows Blom's approximation and $s_b$ is measured from later fixed-parameter evaluations \cite{blom1958statistical}. The identity line is a diagnostic conditional on that scale, not a first-principles prediction. \textbf{b}, $d_{\rm sel}/b$ against $b$ on the same horizontal range; the line marks unity. Points are cell means, with no uncertainty bars shown; their clustered interval is reported separately. Labels identify molecule and backend (A/P); open symbols denote single-source stress cells.}
    \label{fig:dsel-prediction}
\end{figure}

\begin{figure}[H]
    \centering
    \includegraphics[width=0.92\linewidth]{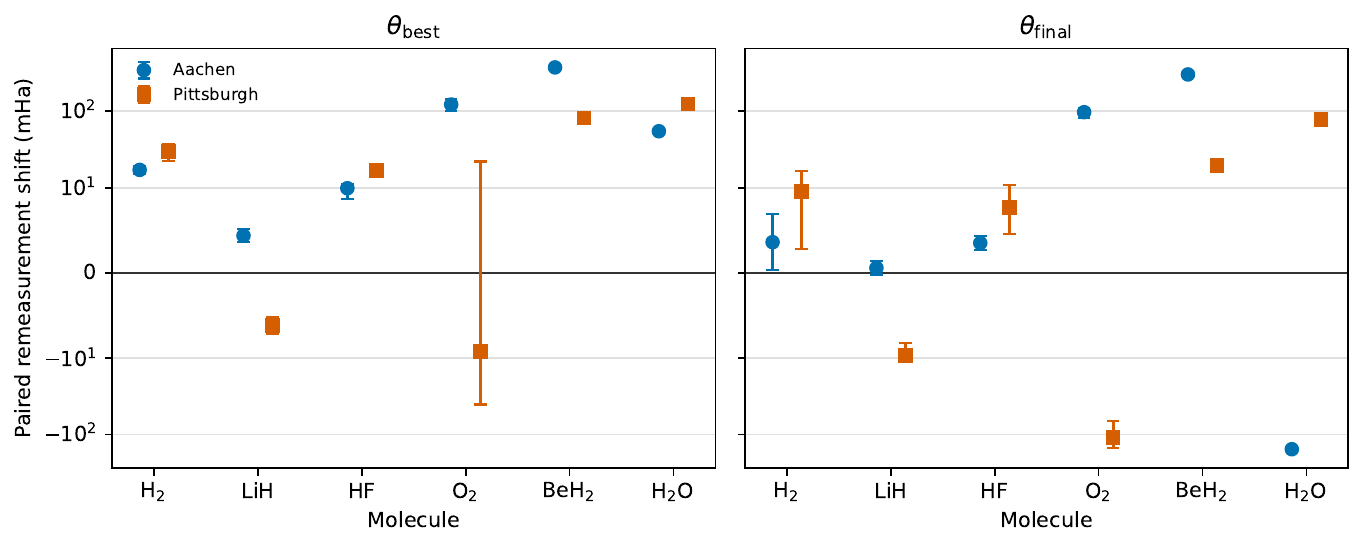}
    \caption{Change in active-space energy when $\theta_\mathrm{best}$ (left) and $\theta_\mathrm{final}$ (right) are remeasured independently without mitigation. Markers show mean paired shifts; error bars are $\pm1$ source-record SEM and are omitted for the single-source BeH$_2$ and H$_2$O stress tests. The symmetric-logarithmic ordinate preserves the small core effects while displaying the full interval from the Pittsburgh O$_2$ lower bar below $-150$ mHa to the Aachen BeH$_2$ shift of $+373$ mHa. Positive values indicate a higher remeasured energy.}
    \label{fig:fixed-validation}
\end{figure}

Figure~\ref{fig:dsel-prediction} summarizes the drift-differenced scale comparison; Table~\ref{tab:fixed-validation} and Figure~\ref{fig:fixed-validation} show molecule- and backend-dependent remeasurement. The cross-cell rank association ($\rho=0.952$) cannot distinguish agreement in magnitude from a shared noise-scale dependence: both $d_{\rm sel}$ and $b$ increase with the measured scatter. The normalized comparison is therefore the ratio $d_{\rm sel}/b=1.014$ and its interval $[\clusterCoreRatioLow{},\clusterCoreRatioHigh{}]$, together with the within-cell centered association reported below. Because $b$ and $d_{\rm sel}$ both use the later fixed-parameter measurements, $d_{\rm sel}/b$ is a scale check rather than an independent test. On \texttt{ibm\_pittsburgh}, H$_2$ and HF shift upward by 29.9 and 16.7 mHa, LiH shifts downward by 6.2 mHa, and O$_2$ is unresolved. On \texttt{ibm\_aachen}, H$_2$, LiH, and HF shift upward by 17.1, 4.4, and 9.9 mHa; all three survive Holm correction over the four molecular tests, while the 122.2 mHa O$_2$ shift remains unresolved at $n=2$ (Table~\ref{tab:fixed-validation}).
\subsection*{Drift-differenced decomposition}
\begin{figure}[H]
\centering
\includegraphics[width=0.78\linewidth]{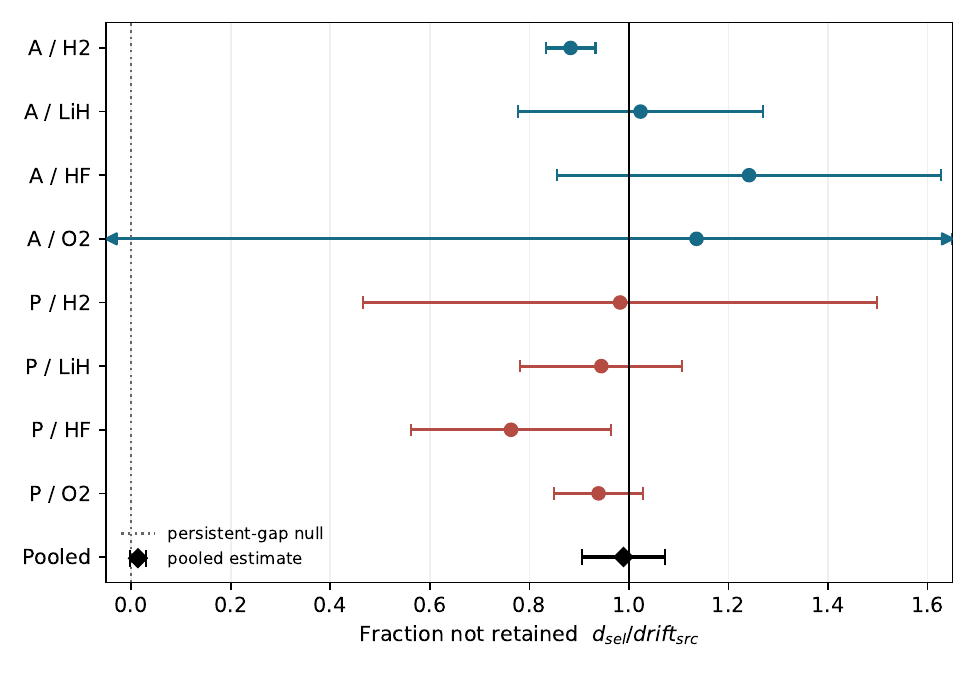}
\caption{Fraction of the source-observed best--final advantage not retained on later fixed-parameter measurement for the eight replicated cells. Markers use the same ratio-of-means estimator as Table~\ref{tab:dsel-primary}; unity means that the apparent source advantage disappears. Cell bars are approximate delta-method intervals with a $t_{n-1}$ critical value, and arrows mark truncated intervals. The pooled diamond uses the cell-balanced session-cluster interval. A/P denote Aachen/Pittsburgh.}
\label{fig:recovery-forest}
\end{figure}

Figure~\ref{fig:recovery-forest} and Table~\ref{tab:dsel-primary} report the paired contrast $d_\mathrm{sel}=\mathrm{drift}_\mathrm{src}+\Delta_\mathrm{fix}$. The persistent-gap null is $\Delta_\mathrm{fix}=-\mathrm{drift}_\mathrm{src}$: the source best--final advantage would remain on remeasurement and $d_{\rm sel}=0$. Mean $\Delta_\mathrm{fix}$ is \coreDeltaFix{} mHa; the cell-balanced non-retained fraction $f_{\rm nr}=d_\mathrm{sel}/\mathrm{drift}_\mathrm{src}$ is \coreRecovery{} ($[\clusterCoreRecoveryLow{},\clusterCoreRecoveryHigh{}]$). A value of one means that none of the source-observed advantage remains at remeasurement. The ratio of sums gives \coreRecoveryRatioOfMeans{} and inverse-variance weighting gives \coreRecoveryPrecisionWeighted{}; Aachen H$_2$ receives 85.3\% of the latter weight. Exact two-sided session-sign enumeration gives $p=0.0854$ for the replicated core. This $p$ tests how unusual the concordant signs of the eight cell-mean $d_{\rm sel}$ values are under session sign symmetry, whereas $[\clusterCoreRecoveryLow{},\clusterCoreRecoveryHigh{}]$ is an approximate interval for the cell-balanced $f_{\rm nr}$, so the two do not test the same null (Supplementary Table~\ref{tab:cluster-inference}). The later measurements support a small retained gap on average, but cannot isolate selection from parameter-dependent change across epochs.

Leaving out each replicated cell in turn gives cell-balanced non-retained fractions from \coreRecoveryLooLow{} to \coreRecoveryLooHigh{}. Record-level final-point shifts, drift-differenced contrasts and cross-backend comparisons are in Supplementary Table~\ref{tab:theta-final-control}.

A second fixed-parameter campaign repeated both saved points after \epochLagMin{}--\epochLagMax{}\,d (Supplementary Table~\ref{tab:epoch-remeasurement}). Across the replicated cells, the mean absolute change in cell-mean $d_{\rm sel}$ was \epochMeanAbsCellDselChange{} mHa and the largest was \epochMaxCellDselChange{} mHa, compared with a primary mean source gap of \epochPrimaryDrift{} mHa. The largest change was \epochMaxChangePctDrift{}\% of that gap. The propagated one-standard-deviation resolutions of the cell changes ranged from \epochResolutionRange{} mHa; no observed change exceeded \epochMaxChangeOverResolution{} times its cell's resolution. Identical best and final points individually moved by as much as \epochMaxBestShift{} and \epochMaxFinalShift{} mHa, respectively. These are observed same-stack changes at the sampled epochs, not a bound on changes between the earlier source and fixed acquisitions.

\subsection*{Additional controls}
The Pittsburgh H$_2$ zero-source-gap record is excluded from the selection endpoint because its denominator is zero. Its distinct parameter vectors have identical saved source energies; a software artifact cannot be excluded without raw shot outcomes or a complete binding audit. The individual diagnostic and the single-source deep-circuit results are in the Supplementary Information.
\subsection*{Order-statistic scaling}
The measured within-vector scatter provides the empirical scale in $b=a_m\sigma$. Cell medians span \sigmaMin{}--\sigmaMax{} mHa, so the raw cross-cell association is expected to be scale dominated. Static calibration does not explain the ordering: Pittsburgh has larger matched-molecule scatter in five of six cells despite its lower device-wide median two-qubit error. Source-plateau and later fixed-vector scatter agree within a factor of 0.63--1.40 in ten of 12 cells, but the coefficient-norm proxy overpredicts scatter by 2.2--6.3 times for the compact tapered Hamiltonians (Supplementary Table~\ref{tab:noise-proxy-ratios} and Figure~\ref{fig:selection-noise}).

Keeping the later fixed-point scatter unchanged, using only final-plateau evaluations as competitors raises cell-level $d_{\rm sel}/b$ by \competitorCellGainMin{}--\competitorCellGainMax{}\%; the per-record range is \competitorRunGainMin{}--\competitorRunGainMax{}\% (Supplementary Table~\ref{tab:competitor-count}). Exploratory within-cell regressions on lag, calibration changes and parameter distance are reported in Supplementary Table~\ref{tab:exploratory-associations}; their sparse within-cell variation limits interpretation.

An offline quadratic-plus-Gaussian experiment with actual COBYLA and SPSA trajectories shows negative selected-observation bias across one to 16 parameters and budgets of 20--500 evaluations (Supplementary Table~\ref{tab:synthetic-orderstat}). This demonstrates that finite-shot selection can persist beyond one parameter in a stylized landscape; it neither models these devices nor removes the need for a higher-dimensional hardware control. At the measured precision, a nominal fixed-point SEM below half the chemical-accuracy threshold would require \precisionHtwoCount{} evaluations for H$_2$ and \precisionWaterCount{} for H$_2$O (Supplementary Table~\ref{tab:noise-floor-requirements}). These are precision-planning counts, not evidence of accuracy.
\subsection*{Mitigation sensitivity}
In Runtime 0.49.0, level 1 enables measurement-error mitigation, whereas level 2 additionally enables zero-noise extrapolation by default. At level 1, the H$_2$ mean lies below CASCI while the HF mean error falls from 21.71 to 6.46 mHa. O$_2$ has a non-monotonic response: its mean error increases at level 1 and decreases at level 2. These separately acquired measurements do not isolate each mitigation component from acquisition changes.
The resilience-1 H$_2$ best-vector mean lies 3.33 mHa below CASCI (Supplementary Table~\ref{tab:resilience-comparison}). Mitigated estimates need not obey the variational upper bound. Because mitigation was applied to fixed vectors without a new optimization, this observation concerns estimator accuracy and does not measure selection bias in a mitigated trajectory.
Level 2 improves H$_2$, LiH, HF and O$_2$ relative to level 0, particularly O$_2$, but does not establish chemical accuracy because finite-shot variation remains large. For H$_2$, the level-2 mean error is 1.20 mHa with a source-vector-level 95\% CI from $-11.65$ to 14.05 mHa, and is not statistically resolved as chemically accurate.
\subsection*{Post hoc Phoenix observation}
Five Phoenix LiH runs form an inconclusive cross-family observation. Their fixed-job and source-plateau noise scales yield different $d_{\rm sel}/b$ conclusions, while processor family, calibration, Runtime stack and transpiler seed differ together. They are not pooled with the Heron endpoint; details are in Supplementary Table~\ref{tab:phoenix-control}.

\section*{Discussion}

Later fixed-parameter measurement does not reproduce the source best--final gap. The non-retained fraction obeys $f_{\rm nr}=d_\mathrm{sel}/\mathrm{drift}_\mathrm{src}=1+\Delta_\mathrm{fix}/\mathrm{drift}_\mathrm{src}$; unity means the apparent advantage disappears. Its cell-balanced estimate is \coreRecovery{}, with ratio-of-sums and inverse-variance estimates of \coreRecoveryRatioOfMeans{} and \coreRecoveryPrecisionWeighted{}. The result is compatible with finite-shot selection on the measured scale, but the separation of acquisition epochs prevents pure causal attribution. The singleton stress cells do not strengthen the core inference.

The second measurement epoch is consistent with common-mode cancellation over the sampled interval: raw fixed-point energies changed appreciably, but paired best and final changes largely tracked, leaving cell-mean $d_\mathrm{sel}$ changes within \epochMaxCellDselChange{} mHa across the \epochCellsWord{} completed cells. Parameter-dependent changes between source and fixed acquisitions can still survive.

The direction of the raw best-point shift is not stable across backends. LiH moves by +4.37 mHa on Aachen and $-6.20$ mHa on Pittsburgh, yet $d_\mathrm{sel}$ is +3.79 and +3.52 mHa. Aachen HF provides the clearest reporting example: a run with a best-observed error of $+0.069$ mHa remeasures at $+9.314$ mHa. Thus neither the best nor final trajectory value alone supports an accuracy claim.

Mitigation improves some fixed-point means but does not establish chemical accuracy at the observed precision. No cell passes equivalence testing against $[-1.594,+1.594]$ mHa (Supplementary Table~\ref{tab:accuracy-tost}). With only $n=2$--5 source records per replicated cell, the intervals are too wide to establish such equivalence. A failed test is therefore not evidence of inaccuracy or of a zero effect.

These results suggest six reporting priorities for hardware VQE: (1) remeasure claimed parameter points within the optimization session when feasible; (2) publish the complete trajectory and distinguish best-observed from final values; (3) archive shot allocation, resolved Runtime options and per-evaluation outcomes; (4) report independent source-run counts, uncertainty and acquisition lag; (5) preserve circuit, layout and used-resource calibration provenance; and (6) test chemical accuracy as equivalence to a stated threshold, rather than selecting a lucky minimum.

\subsection*{Interpretation and fixed-epoch correction}
The pooled non-retained fraction does not establish energetic equivalence separately in every cell. In LiH, HF and O$_2$, the measured contrast can exceed the active-space correlation signal. The same quantity in the single-source deep-circuit cells is descriptive, not a replicated estimate.

The practical fixed-epoch correction $E_{\rm best}+a_m\hat\sigma$ leaves Aachen residual gaps of 0.66, 2.74, and 6.40 mHa for LiH, HF, and H$_2$, removing 63--85\% of their remeasurement gaps. Across the eight core cells it under-corrects \debiasedUnderCount{} and over-corrects \debiasedOverCount{}; Pittsburgh LiH is over-corrected because epoch drift dominates. The formula is a diagnostic and does not replace remeasurement.

\subsection*{Limitations}
The principal limitation is the absence of a same-session control: source and fixed measurements are separated by 5.6--38.4 days. Saved local package-version stamps also differ between most source records and the fixed acquisitions. A paired offline comparison finds numerical agreement of reconstructed states, energies and compact noisy controls across those SDK versions, but does not replay historical remote execution or resolved server options. Differencing cancels a change common to both parameter points, but not a parameter-dependent one. A same-session comparison is the necessary next control.

The replicated mechanism is tested only in near-minimal circuits: all eight cells have one variational parameter, one to four logical qubits, at most 12 two-qubit gates and roughly 18--27 evaluations. Extrapolation to high-dimensional VQE is untested. BeH$_2$ and H$_2$O have one source run per backend and different optimizer budgets; only H$_2$ has active-space correlation above the chemical-accuracy threshold among the replicated molecules.
The interval describes the selected benchmark cells conditional on their recorded campaigns. Source-session clustering does not estimate additional uncertainty from correlations between later fixed acquisitions, systematic epoch effects, or a broader population of molecules and devices.

The structured archive lacks raw per-evaluation shot outcomes and per-term counts, precluding retrospective shot splitting. It also lacks a verbatim dump of resolved source Runtime options, shot-allocation semantics and post-grouping circuit counts. Only one equilibrium geometry per molecule, STO-3G active spaces and unmitigated primary trajectories were studied.

The contrast cancels only calibration changes common to both vectors. The magnitude of the raw best-point shift exceeds the first-order readout envelope in eight of 12 cells; $|d_\mathrm{sel}|$ exceeds it in six of 12 (Supplementary Table~\ref{tab:readout-drift-envelope}). The empirical order-statistic scale uses the full archived trajectory length because every evaluation could compete for the reported minimum; this primary choice was not preregistered. Restricting the effective competitor count to the final-plateau length raises $d_\mathrm{sel}/b$ by 31.9--56.4\%, and both definitions are reported as sensitivity analyses. The coefficient-norm proxy cannot replace the measured scale because it overpredicts scatter by 2.2--6.3 times for the tapered one- and two-qubit Hamiltonians.

The core uses one processor family, although Aachen and Pittsburgh are distinct instruments in different regions. In the operator-supplied 2026-09-09 fleet snapshot (Supplementary Table~\ref{tab:qpu-inventory}), the selected systems occupy the leading error tier but are not the fleet minima. Nighthawk r1 alternatives combined substantially lower throughput with higher reported errors, so a campaign on them would confound family with error regime. Phoenix availability during the core campaign is undocumented, and its post hoc LiH control is not pooled. Device availability and calibration are time dependent.

\subsection*{Conclusions and outlook}
This work examines apparent finite-shot selection in hardware VQE using a molecule ladder chosen for hardware tractability. In the replicated cells, almost all of the optimizer-selected best--final advantage is absent on later fixed-parameter measurement, at a scale consistent with an empirical order-statistic model. Within the replicated set, H$_2$ is the only correlation-recovery case above the chemical-accuracy threshold; LiH, HF, and symmetry-broken O$_2$ primarily probe estimator reproducibility, while BeH$_2$ and H$_2$O are deep-circuit stress tests.
The deeper BeH$_2$ and H$_2$O circuits remain far from their active-space references, and compact noisy Aer models do not reliably rank devices. The transferable result is therefore methodological: publish full trajectories, confirm selected parameter points independently and promptly, preserve circuit and calibration provenance, and test accuracy through equivalence rather than a lucky minimum. A same-session experiment on higher-dimensional ansatzes is the next necessary test of generality.

\section*{Methods}

Cartesian molecular geometries were specified in \AA{} and used to build second-quantized electronic Hamiltonians in the STO-3G basis. H$_2$O has two 0.9685 \AA{} O--H bonds and a 103.6$^\circ$ angle; BeH$_2$ has 1.3295 \AA{} Be--H bonds. The archived BeH$_2$ geometry includes a 0.001303 \AA{} transverse coordinate, corresponding to 0.06$^\circ$ from linearity; the recorded coordinates were used without imposing exact linear symmetry.
The active spaces were fixed before hardware execution.
H$_2$, LiH, and HF used two active electrons in two spatial orbitals (2e2o).
O$_2$, BeH$_2$, and H$_2$O used four active electrons in four spatial orbitals (4e4o).
In Qiskit Nature this corresponds to applying an \texttt{ActiveSpaceTransformer} to the PySCF-derived electronic-structure problem before fermion-to-qubit mapping.
The reference energy for each molecule is the lowest eigenvalue in the corresponding particle-number and parity-reduced tapered sector.
Independent diagonalization before tapering confirmed that all six selected minima have the intended particle number and total spin and agree with the corresponding tapered eigenvalues to better than $10^{-12}$ Ha.
These checks use the active electronic Hamiltonian and exclude nuclear-repulsion and inactive-space constants.

The fermionic Hamiltonians were mapped using the parity transformation with particle-number tapering (PT) and subsequent symmetry tapering.
The variational state was a UCC ansatz initialized from the Hartree--Fock state.
O$_2$ was treated as the triplet $^3\Sigma_g^-$ state with active populations $(N_\alpha,N_\beta)=(3,1)$.
The keyword 4e4o specification takes precedence over the advisory YAML indices, and Qiskit Nature's zero-indexed automatic selection used spatial orbitals 6--9. These indices were reconstructed post hoc from the ROHF audit and were not materialized in the archived execution records.
The reconstructed $D_{\infty h}$ ROHF orbitals retain the degenerate $\pi_g^*$ pair as zero-indexed orbitals 7 and 8 within the 6--9 window (Supplementary Table~\ref{tab:o2-symmetry}) but include only the E1ux component of the bonding $\pi_u$ pair. The compact active space therefore breaks $\pi_u$ degeneracy, so its CASCI value is a symmetry-broken active-space reference; the fixed triplet sector and symmetry restrictions reduce the retained UCC doubles manifold to one independent amplitude.
The physical and tapered minima both have $N=4$ and $\langle S^2\rangle=2$, with an energy difference below $10^{-12}$ Ha.
A unique state-vector overlap is not reported because tapering changes the Hilbert-space dimension and the numerical ground vector may mix degenerate spin projections within the triplet manifold.
After particle-number reduction, the 4e4o parity-mapped register contains six qubits. The tapered mapper removes two additional $Z_2$ symmetries for O$_2$, leaving four qubits, but removes one for BeH$_2$ and H$_2$O, leaving five. Particle number, spin population, parity, and the truncated UCC doubles excitation manifold leave one independent O$_2$ amplitude. The optimized noiseless UCC value reproduces the active-space CASCI eigenvalue to six decimal places, confirming exact expressibility within this restricted sector rather than for the unrestricted 4e4o Hilbert space.

Supplementary Table~\ref{tab:energy-conventions} reports both Hamiltonian conventions used for reproducibility.
For any hardware or simulator value, the corresponding total molecular energy is obtained by adding $E_{\rm nuc}+E_{\rm inactive}$; because this offset is common to the estimate and its CASCI reference, the signed errors reported below are unchanged.
Circuits were transpiled with \texttt{optimization\_level=0}, which suppresses optional circuit-optimization passes while retaining target-specific basis translation and routing.
This makes the transpilation choice transparent, but level 0 still performs target-specific routing; the circuit depths and two-qubit gate counts in \Cref{tab:resources} therefore remain backend dependent.
The resource counts therefore describe the archived level-0 executions; they are not minimum achievable costs for these ansatzes.
Energy expectation values were obtained with the IBM Runtime estimator.
The configured target estimator precision was \(1/\sqrt{1024}=0.03125\), corresponding to the nominal 1024-shot convention in the Runtime primitive interface.
Saved Runtime metadata do not determine whether the 1024 shots were allocated per evaluation, Pauli term, or commuting group, so no physical per-evaluation variance bound is inferred from the archived records.
Instead, Supplementary Table~\ref{tab:hardware-plateau-sigma} reports per-run standard deviations over the final 30\% of each trajectory, together with identity-excluded Hamiltonian coefficient-norm proxies at the nominal shot count.
The proxies provide a common scale for the active-space operators, but covariance from grouped measurements prevents them from establishing whether shot noise or optimizer dynamics dominates an individual plateau.
Classical parameters were optimized with COBYLA from the same all-zero parameter vector in every run; no random initialization or initialization seed was used. The Qiskit optimizer defaults were retained, with \texttt{rhobeg=1.0} and \texttt{tol=None}. The saved configuration used \texttt{maxiter=40} for H$_2$, LiH, HF, O$_2$, and the \texttt{ibm\_aachen} BeH$_2$/H$_2$O runs, whereas the \texttt{ibm\_pittsburgh} BeH$_2$/H$_2$O runs used \texttt{maxiter=30}; \texttt{maxiter} is an upper bound, and most low-dimensional runs terminated after 18--27 evaluations under COBYLA's default stopping rules. Exact termination messages were not archived; actual evaluation counts are listed in Supplementary Table~\ref{tab:hardware-plateau-sigma}.

The VQE optimization trajectories and primary fixed-parameter measurements used the archived resilience-level-0 configuration. No zero-noise extrapolation or measurement-error mitigation was requested. Complete resolved source options were not saved, so the archive cannot independently verify every execution default, including dynamical decoupling and suppression options.
A smaller fixed-parameter mitigation-sensitivity dataset was then collected on \texttt{ibm\_pittsburgh} using Runtime resilience levels 1 and 2 on the same molecule and parameter-vector set, with no COBYLA updates; it is analyzed separately and is not folded into the raw hardware benchmark means.

For each hardware run, the full sequence of measured objective values is analyzed.
The reported energy is the best observed value during the optimization trajectory.
This convention is useful for diagnosing whether the optimizer ever sampled a low-energy point, but it can favor statistically lucky finite-shot evaluations; recent VQE work identifies this selection bias as a ``winner's curse'' in noisy variational optimization \cite{novak2025reliable}.
The difference between the final evaluated energy and the best observed energy is used as a simple convergence-drift diagnostic.
The drift-differenced contrast uses later remeasurements of both parameter points to remove calibration shifts common to that epoch. It cannot separate selection from parameter-dependent changes between epochs.
To assess the selection interpretation, saved \(\theta_\mathrm{best}\) and \(\theta_\mathrm{final}\) vectors were remeasured independently, meaning in fresh fixed-parameter Runtime estimator jobs during a later calibration epoch by the same author and pipeline, using the same transpiled circuit construction, layout policy, 1024-shot precision, and no COBYLA updates.
The primary resilience-0 dataset comprises 676 evaluations of 66 vectors from all 33 primary source runs. It combines ten evaluations per vector for the 17-run Pittsburgh core (340 evaluations), ten per vector for the 12-run Aachen core (240 evaluations), and 12 per vector for the two stress-test source runs on each backend (48 evaluations per backend; 96 total). A second campaign added \epochEvaluations{} core evaluations (\epochAachenEvaluations{} Aachen and \epochPittsburghEvaluations{} Pittsburgh), including the successful split Pittsburgh O$_2$ retry; the original timed-out job remains archived as a failure. Thus \allResilienceZeroEvaluations{} successful resilience-0 evaluations are archived, while only the original 676 enter the primary selection analysis. Evaluation-level scatter combines shot noise with vector-to-vector variation; source-run means are used as the independent units for paired inference.
Multiplicity adjustment is applied separately within each backend, and source runs are not pooled across backends. The primary non-retained-fraction endpoint excludes the Pittsburgh H$_2$ zero-source-gap record, so that cell has $n=3$ in contrast analyses. Raw best-point shifts use all four independently remeasured records and are a separate, $n=4$ descriptive endpoint; the two sample sizes must not be interchanged.
\subsection*{Statistical analysis}
The historical symbol $\mathrm{drift}_{\rm src}$ denotes the source final-minus-best gap; it is not an estimate of calibration drift during optimization.
Blom's scale describes the minimum of independent, equal-variance Gaussian observations at a common true energy. Adaptive optimizer trajectories need not satisfy those assumptions: their true energies, variances and temporal dependence can change, and the source minimum also influences the final parameter point. Using the full trajectory length and later measured scatter therefore supplies an empirical scale comparison rather than an unbiased estimate of the source selection effect.

\textbf{Analysis choices.} The archived execution configuration identifies the molecule, circuit and best-observed reporting convention for each run. No formal preregistration establishes the cross-cell endpoint or analysis choices. The cell-balanced non-retained fraction, full-trajectory competitor count and $[0.8,1.2]$ margin were selected during analysis. The Phoenix decision rule lacks independently verifiable pre-analysis timing and is not used as prospective evidence. All inferential comparisons are exploratory.

Supporting within-cell analyses standardize each predictor and outcome within backend--molecule cells, omit cells without within-cell variation, and obtain two-sided exploratory $p$ values by \exploratoryPermutations{} seeded permutations of the predictor within cells. These permutations do not preserve session clustering and are not confirmatory tests.

For each source record $i$, we define $d_{\mathrm{sel},i}=(E^\mathrm{fix}_\mathrm{best}-E^\mathrm{src}_\mathrm{best})-(E^\mathrm{fix}_\mathrm{final}-E^\mathrm{src}_\mathrm{final})$. With $\mathrm{drift}_{\mathrm{src},i}=E^\mathrm{src}_\mathrm{final}-E^\mathrm{src}_\mathrm{best}$ and $\Delta_{\mathrm{fix},i}=E^\mathrm{fix}_\mathrm{best}-E^\mathrm{fix}_\mathrm{final}$, the identity $d_{\mathrm{sel},i}=\mathrm{drift}_{\mathrm{src},i}+\Delta_{\mathrm{fix},i}$ holds. The persistent-gap null $\Delta_{\mathrm{fix},i}=-\mathrm{drift}_{\mathrm{src},i}$ gives $d_{\mathrm{sel},i}=0$. The non-retained fraction is $f_{\rm nr}=d_{\mathrm{sel}}/\mathrm{drift}_{\mathrm{src}}$; archived CSVs retain the historical column name \texttt{recovery\_fraction} for compatibility. Both fixed-parameter means were acquired in the same later campaign, so the difference removes an additive epoch shift common to the two points to first order, but not parameter-dependent changes between epochs. Let $s_{b,i}$ be the sample SD across repeated fixed measurements of $\theta_{\rm best}$ and $a_{m_i}=\Phi^{-1}[(m_i-0.375)/(m_i+0.25)]$, where $m_i$ is source trajectory length. The run-level Gaussian diagnostic is $b_i=a_{m_i}s_{b,i}$, averaged within a cell; it is not the product of cell medians \cite{blom1958statistical}. Supplementary Table~\ref{tab:paired-remeasurement} publishes its factors. Source-plateau and pooled best/final SDs support sensitivity and precision planning, not primary hypothesis tests.

For cell $c$, the endpoint uses the ratio of record means $f_{{\rm nr},c}=\bar d_{{\rm sel},c}/\overline{\mathrm{drift}}_{{\rm src},c}$; the primary estimate is the equally weighted mean $C^{-1}\sum_c f_{{\rm nr},c}$ across the replicated cells. It is neither a pooled ratio nor a mean of individual-record ratios. The quantity is unbounded: values above one indicate reversal of the mean best--final ordering, and negative values indicate a larger retained advantage. It is not a probability or the proportion of records showing recovery.

The 95\% enumerated source-session wild-cluster Rademacher interval conditions on those cells and sign-flips linearized record influences over complete source sessions. Enumeration removes Monte Carlo error, not finite-sample approximation: the interval uses estimated influences and an approximate sign-symmetric cluster distribution. Source records are the observational units; dependence among records sharing a source session is handled by clustering, and the eight cells span 15 sessions. The session-sign $p$ value also assumes sign symmetry and is not a randomization test from an assigned intervention. Within-cell paired tests use distinct source sessions, and cross-backend tests use separate Holm families. Further interval and permutation details are in Supplementary Table~\ref{tab:cluster-inference}.

Pittsburgh core record-level lags were 15.6--32.0 d (cell medians 28.8--29.9 d; Supplementary Table~\ref{tab:remeasurement-lag}); Aachen core lags were approximately 37 d, Pittsburgh stress-test lags about 6 d, and Aachen stress-test lags 38.2--38.4 d. These values use source-run identifiers and fixed-job creation times because exact source-job completion times are absent from the clean index.

Remeasurements were performed in later epochs because they were added after the archived optimizations; a same-session, zero-lag remeasurement is the designated next control. The primary remeasurement used resilience level 0, while the follow-up mitigation-sensitivity measurements used resilience levels 1 and 2 on the same molecule and parameter-vector set.
Each table and figure identifies its uncertainty definition and sampling unit: within-vector SD, source-record SEM or t interval, or session-cluster interval. Single-source cells do not have an independently estimated between-source uncertainty.

Exact duplicate files and records without successful completion status were excluded from the quantitative summaries. The Phoenix ledger also retains two completed jobs from a superseded acquisition, but their evaluation payloads were overwritten when its run identifier was reused. Quantitative analysis therefore uses the canonical acquisition with retained payloads; this is a data-availability exclusion, not an exact-duplicate claim.

Simulator baselines used the same active-space Hamiltonians, tapered-parity mapping, and UCC ansatz. The call \texttt{ansatz.decompose(reps=10)} sets recursive decomposition depth rather than ansatz repetitions, so hardware and simulator circuits are logically identical.
The ideal baseline used noiseless statevector expectation values. The corrected noisy baselines evaluate every archived $\theta_\mathrm{best}$ vector with an exact Qiskit Aer density-matrix expectation under compact depolarizing models whose one- and two-qubit error probabilities were estimated from the recorded backend calibration aggregates. They contain neither the EstimatorV2 precision Gaussian nor best-of-trajectory selection.
Fixed-parameter mock-backend simulations used \texttt{AerSimulator.from\_backend} with local IBM Runtime fake-backend snapshots. Their workflow checks and the excluded cloud-service diagnostic are documented in the Supplementary Information.

Both core-campaign backends were 156-qubit Heron r3 processors with native \texttt{cz}, \texttt{id}, \texttt{rz}, \texttt{sx}, and \texttt{x} operations (Supplementary Table~\ref{tab:qpu-inventory}).
Source calibration summaries use the saved initial-layout qubits and directed coupler entries rather than device-wide averages. Full post-routing touched-qubit sets and complete fixed-epoch layouts were not retained, so these summaries do not certify calibration coverage of every executed resource.
The eight-cell core design was fixed and executing before \texttt{ibm\_phoenix} appeared in the first instance-list snapshot retained by this project. That snapshot, dated 2026-09-03, postdates the core source campaign; IBM announced Pay-As-You-Go access to the Nighthawk r2 device in September 2026 without specifying an exact day \cite{ibm2026compute-changelog}. Operational availability during the core campaign is not asserted. One LiH source run submitted on 2026-09-09 and four on 2026-09-10 form a dated, post hoc, single-cell cross-family control. They are reported separately and are never pooled into the primary core or its clustered inference. Source-run transpiler seeds are not preserved; the fixed remeasurement used seed 1234.

The reporting protocol used here separates molecular definition, circuit construction, noisy optimization, independent remeasurement, and exact active-space reference evaluation.
Each molecule is defined by a Cartesian geometry, charge, spin, STO-3G basis, and active-space specification before any hardware call is made.
For every run, the analysis preserves four distinct energy estimates: the best observed finite-shot optimizer value, the final optimizer value, the noiseless UCC value at saved parameters, and the exact CASCI eigenvalue of the same active-space qubit Hamiltonian.
For all six molecules, a fifth estimate is obtained by independently remeasuring saved \(\theta_\mathrm{best}\) and \(\theta_\mathrm{final}\) values in fixed-parameter estimator jobs; the four-molecule core and the two single-source stress tests span both backends.
This separation is essential because these quantities answer different questions: reachability of low-energy samples, optimizer termination quality, ansatz expressivity, active-space reference accuracy, and finite-shot selection bias.

The acquisition record contains molecular inputs, active-space definition, mapper and tapering choice, ansatz settings, logical and transpiled circuit resources, physical layout, used-qubit calibration summaries, optimization trajectories, Runtime job identifiers, and usage metadata. Provider job identifiers are not part of the public deposit.
Completed-job billing and the source/fixed usage ledger are documented in Supplementary Table~\ref{tab:qpu-usage}; queue and wall-clock time are excluded.
This makes the dataset a baseline against which future mitigated, adaptive-ansatz, or larger-active-space studies can be compared directly.

For the primary dataset, the cached result index records 14 successful, 1 failed, 1 running, and 1 missing-status Aachen entries, and 21 successful, 1 failed, 3 running, and 1 failed-recovered Pittsburgh entries; two exact-duplicate Pittsburgh records are excluded, leaving 33 molecule-level records from 17 distinct hardware sessions. Phoenix is indexed separately as five successful source records; its canonical fixed acquisition contains 100 successful evaluations, and the job ledger retains two additional completed jobs from the superseded duplicate attempt. Within every primary molecule/backend cell, records come from distinct sessions; Aachen H$_2$, LiH, and HF share three acquisition sessions. Session keys are assigned chronologically within each backend; raw run identifiers retain the period- and hyphen-separated date formats present in the archive. The current analysis environment reports Qiskit 2.5.2, Qiskit Nature 0.7.2, Qiskit Aer 0.17.2, Qiskit IBM Runtime 0.49.0, and PySCF 2.10.0. Archived source metadata record resilience level 0 and precision 0.03125, but do not preserve a complete verbatim dump of those options. Saved local package stamps and separate remote job labels are summarized in Supplementary Table~\ref{tab:software-provenance} and linked per record in \texttt{software\_provenance.json}, supplied with the dataset. Supplementary Table~\ref{tab:software-compatibility} reports a paired local SDK comparison at the published parameter vectors, with common classical dependencies held fixed.
Supplementary Tables~\ref{tab:geometry-active-space}, \ref{tab:backends}, \ref{tab:hardware-plateau-sigma}, and \ref{tab:run-details} provide molecular inputs and energy offsets, backend characteristics, plateau statistics, and source-run details.
\section*{Data availability}
\newcommand{\dataDOI}{https://doi.org/10.5281/zenodo.23082359}

A curated archive of the data supporting this study is publicly available at \url{\dataDOI} \cite{larrucea2026beyondh2data}. It includes the primary and post hoc Phoenix records, optimization evaluations, fixed-parameter measurements, circuit and calibration metadata, reference energies, simulation controls, schemas and an integrity manifest. The data are licensed under CC BY 4.0.

Zenodo is the archival copy; the MolQBench web application at \url{https://quantum.larrucea.eu/molqbench/} is a searchable convenience mirror.

\section*{Code availability}

The Zenodo dataset at \url{\dataDOI} includes MIT-licensed acquisition examples and an offline analysis package. The package reproduces the principal tables and inference from the archived CSVs without provider credentials. Separate scripts generate the fixed-seed stylized optimizer experiment and the paired local software-compatibility controls; \texttt{analysis/COVERAGE.md} identifies supplementary results and figures not regenerated by this package.

\bibliographystyle{nature5}
\bibliography{references}

\section*{Acknowledgements}

The author gratefully acknowledges the Fraunhofer internal program ``Quantum Now'' (Editions 5 and 6) for providing access to the IBM Quantum platform and the quantum-computing resources used in this work.

This work was carried out within the project ``Evaluierung von Quantencomputerressourcen f\"ur elektronische Strukturberechnungen (EQeS), Teilvorhaben: Chemie und Analyse auf Anwendungsebene'', funded by the German Federal Ministry of Research, Technology and Space (Bundesministerium f\"ur Forschung, Technologie und Raumfahrt, BMFTR) under contract number 13N17338 as part of the Research Program Quantum Systems (Forschungsprogramm Quantensysteme).

\section*{Author contributions}

J.L. was solely responsible for conceptualization, funding acquisition, methodology, software, formal analysis, investigation and execution of the calculations, validation, resources, data curation, visualization, project administration, and writing and revision of the manuscript.

\section*{Competing interests}

The author declares no competing interests.

\clearpage
\section*{Supplementary Information}
\setcounter{table}{0}
\renewcommand{\thetable}{S\arabic{table}}
\renewcommand{\theHtable}{S\arabic{table}}
\setcounter{figure}{0}
\renewcommand{\thefigure}{S\arabic{figure}}
\renewcommand{\theHfigure}{suppfigure.\arabic{figure}}

Supplementary Tables provide molecular inputs, energy conventions, prior-work scope, backend metadata, simulator values, Runtime usage, and optimization statistics supporting the main analysis.
\subsection*{Record and control details}
Four retained source records have bitwise-identical saved energies at distinct parameter vectors: Pittsburgh H$_2$ 2026-08-01-12.09.37 (evaluation indices 13/17), Aachen HF 2026-08-01-16.36.57 (12/19), Pittsburgh H$_2$ 2026-08-01-17.54.22 (6/13), and Aachen H$_2$ 2026-08-01-18.57.40 (13/15). Only the first tie joins that record's selected best and final evaluations. Those vectors differ by $\lVert\Delta\theta\rVert_2=1.0\times10^{-3}$; their archived job references and submission times differ. Raw shot outcomes and a complete binding audit are unavailable, so neither a discrete-outcome collision nor a source software artifact is excluded. That record has zero source gap and is omitted from ratio endpoints, not treated as a validated experimental null. Independent fixed jobs give its best- and final-point shifts of $+8.411$ and $-1.994$ mHa, or $d_{\rm sel}=+10.405$ mHa. A second fixed epoch gives $+3.047$ mHa; the change is within its propagated resolution.

As an offline diagnostic, the saved H$_2$/HF Hamiltonians were used to enumerate the possible energies under exactly 1024 independent binary outcomes for each non-identity Pauli term. All \tieLatticeMatches{}/\tieLatticeEvaluations{} saved H$_2$/HF source energies, including the four ties, lie on this conditional lattice within $10^{-10}$ Ha. Among \tieLatticePairOpportunities{} within-run evaluation pairs, four exact ties occur in four runs. A seeded stationary-binomial simulation, with per-run outcome probabilities fitted from lattice-inverted aggregate energies, gives $P(\text{at least four tied runs})=\tieConditionalProbability{}$. This makes finite-shot collisions quantitatively plausible under that model, not established as the acquisition mechanism: the actual term/group shot allocation, outcomes, parameter-dependent probabilities and covariance were not archived, and a software artifact cannot be excluded.

The single-source BeH$_2$ and H$_2$O cells have large, backend-dependent raw shifts and are not used for replicated inference. The ideal UCC baseline is close to active-space CASCI for the compact core and within roughly 1--2 mHa for the deep 4e4o stress tests at the saved budget. The noisy Aer controls are fixed-parameter density-matrix expectations, omit the EstimatorV2 precision Gaussian and do not select an optimizer minimum. Their depolarizing probabilities are calibration-aggregate surrogates, not device twins.

The extended 12-cell scale ratio is 1.111. Its session-sandwich interval is $[0.994,1.228]$; the conditional wild-cluster interval $[\clusterExtendedRatioLow{},\clusterExtendedRatioHigh{}]$ excludes unity only because four singleton cells have zero residual influence. Neither interval includes their sampling uncertainty. Exact two-sided session-sign enumeration gives $p=1400/131072=0.0107$ for the extended cohort, versus $p=2800/32768=0.0854$ for the replicated core. These compare different cohorts, not differently standardized versions of one test, and do not strengthen the core inference.

The post hoc Phoenix LiH cell yields $d_{\rm sel}/b=\phoenixDselOverB{}$ as a mean of record ratios and \phoenixRatioOfMeans{} as a ratio of means when $b$ uses later fixed-job scatter. Substituting source-plateau scatter gives \phoenixPlateauDselRatio{} and \phoenixPlateauDselRatioMeans{}. The purported advance timing of a high-contrast decision rule cannot be independently verified, so that rule has no confirmatory status. Processor family, calibration, Runtime stack and transpiler seed are confounded.
\subsection*{Mock-backend workflow checks}
Local mock-backend simulations were workflow and qualitative-noise checks. For the Pittsburgh mock backend, H$_2$, LiH, and HF mean errors are 2.5--6.3 mHa, compared with hardware fixed-parameter errors of 10--26 mHa. Its O$_2$ error is 170.98 mHa versus 383.07 mHa for the corresponding hardware fixed mean; both are outside chemical accuracy. These mock models test construction and qualitative degradation but are too optimistic for quantitative prediction. A dashboard-visible cloud mock service for Pittsburgh was excluded because a sanity check on $|0\rangle$ with $Z$ returned $-0.0176$ instead of a value near $+1$.

\subsection*{Device scope and fleet context}
The operator-supplied 2026-09-09 snapshot is cited for fleet context only. Layered two-qubit error spans factors of \fleetLayerSpan{} across the listed fleet and \classLayerSpan{} within the 156-qubit class; Pittsburgh ranks \pittsburghLayerRank{} and Aachen \aachenLayerRank{} within that class. Fleet and class median-readout spans are \fleetReadoutSpan{} and \classReadoutSpan{}. Phoenix lists \phoenixClopsRatio{} times Pittsburgh's CLOPS, \phoenixMcpsRatio{} times its MCPS, and \phoenixReadoutRatio{} times its median readout error. The UI rounds MCPS to 3.8 kHz, whereas the archived screening snapshot reports 3829 Hz. Region-matched Nighthawk r1 systems list \rOneClopsLow{}--\rOneClopsHigh{} times lower CLOPS, \rOneMcps{} times lower MCPS, \rOneLayerLow{}--\rOneLayerHigh{} times higher layered two-qubit error, and \rOneReadoutLow{}--\rOneReadoutHigh{} times higher median readout error. The core source campaign ran from 2026-07-30 to 2026-09-02; Phoenix availability during it is undocumented.
\begin{table}[H]
\centering
\scriptsize
\textbf{(a) Molecular and active-space inputs}\par\smallskip
\resizebox{\linewidth}{!}{%
\begin{tabular}{llllll}
\toprule
Molecule & Charge & Multiplicity & Basis & Active space & Geometry (\AA) \\
\midrule
H$_2$ & 0 & 1 & sto3g & 2e2o & H (0,0,0); H (0,0,0.7428) \\
LiH & 0 & 1 & sto3g & 2e2o & Li (0,0,0); H (0,0,1.6204) \\
HF & 0 & 1 & sto3g & 2e2o & F (0,0,0); H (0,0,0.9338) \\
O$_2$ & 0 & 3 & sto3g & 4e4o & O (0,0,0); O (0,0,1.2160) \\
BeH$_2$ & 0 & 1 & sto3g & 4e4o & H (0,0,0); Be (0,0,1.329445); H (0.001303,0,2.658946) \\
H$_2$O & 0 & 1 & sto3g & 4e4o & H (0,0,0); O (0,0,0.968544); H (0.941424,0,1.196314) \\
\bottomrule
\end{tabular}}
\par\medskip
\textbf{(b) Energy conventions}\par\smallskip
\begin{tabular}{lrrrrrr}
\toprule
Molecule & Active HF & Active CASCI & $E_{\rm corr}$ (mHa) & $E_{\rm nuc}$ & $E_{\rm inactive}$ & Total CASCI \\
\midrule
H$_2$ & $-1.829015$ & $-1.849662$ & $-20.65$ & 0.712409 & 0.000000 & $-1.137253$ \\
LiH & $-1.052949$ & $-1.053223$ & $-0.27$ & 0.979716 & $-7.787901$ & $-7.861408$ \\
HF & $-1.924827$ & $-1.925584$ & $-0.76$ & 5.100230 & $-101.747616$ & $-98.572971$ \\
O$_2$ & $-5.361900$ & $-5.362649$ & $-0.75$ & 27.851432 & $-170.121940$ & $-147.633158$ \\
BeH$_2$ & $-3.911856$ & $-3.917749$ & $-5.89$ & 3.383299 & $-15.031574$ & $-15.566024$ \\
H$_2$O & $-6.138012$ & $-6.145814$ & $-7.80$ & 9.089246 & $-77.915751$ & $-74.972318$ \\
\bottomrule
\end{tabular}
\caption{Reproducible molecular definitions and reference energies. Geometries are loaded by \texttt{molecules.py}. Energies are in hartree except active-space correlation energy $E_{\rm corr}=1000(E_{\rm CASCI}-E_{\rm HF})$, reported in mHa. Active HF and CASCI use the transformed Hamiltonian measured by the quantum workflow; adding $E_{\rm nuc}+E_{\rm inactive}$ gives the total convention without changing signed errors.}
\label{tab:geometry-active-space}\label{tab:energy-conventions}
\end{table}

\FloatBarrier
\begin{table}[H]
\centering
\small
\resizebox{\linewidth}{!}{%
\begin{tabular}{llll}
\toprule
Study & Molecular scope & Device & Independent remeasurement \\
\midrule
Kandala \textit{et al.} \cite{kandala2017hardware} & BeH$_2$ and other small molecules & IBM superconducting & not reported \\
O'Malley \textit{et al.} \cite{omalley2016scalable} & H$_2$ & Superconducting & not reported \\
McCaskey \textit{et al.} \cite{mccaskey2019benchmark} & Multiple molecules & NISQ devices & not reported \\
Larrucea \textit{et al.} \cite{julen2026h2bench} & H$_2$ & IBM Quantum hardware & not reported \\
This work & Six molecules & Two Heron r3 processors plus a post hoc LiH control & yes; all six on Heron, LiH on Nighthawk r2 \\
\bottomrule
\end{tabular}
}
\caption{Scope comparison with representative hardware-VQE studies. ``Not reported'' refers specifically to independent fixed-parameter remeasurement in the cited report. The distinguishing feature of this work is molecule-resolved remeasurement of optimizer-selected points.}
\label{tab:prior-work}
\end{table}

\FloatBarrier
\begin{table}[H]
\centering
\scriptsize
\resizebox{\linewidth}{!}{%
\begin{tabular}{lllllllll}
\toprule
Mol. & Backend & Layout q. & Directed edge entries & RO err. (\%) & $T_1$ ($\mu$s) & $T_2$ ($\mu$s) & 2q err. (\%) & 2q dur. (ns) \\
\midrule
H$_2$ & aachen & 1 & 0 & 0.317 & 238.4 & 370.5 &  &  \\
H$_2$ & pittsburgh & 1 & 0 & 1.001 & 212.0 & 252.1 &  &  \\
LiH & aachen & 2 & 1 & 0.464 & 216.2 & 313.7 & 0.079 & 68.0 \\
LiH & pittsburgh & 2 & 1 & 0.696 & 276.2 & 277.7 & 0.226 & 88.0 \\
HF & aachen & 1 & 0 & 0.317 & 238.4 & 370.5 &  &  \\
HF & pittsburgh & 1 & 0 & 1.001 & 212.0 & 252.1 &  &  \\
O$_2$ & aachen & 4 & 3 & 0.763 & 223.7 & 292.9 & 0.215 & 68.0 \\
O$_2$ & pittsburgh & 4 & 3 & 0.854 & 242.5 & 224.1 & 0.369 & 88.0 \\
BeH$_2$ & aachen & 5 & 7 & 1.243 & 219.4 & 307.6 & 0.339 & 68.0 \\
BeH$_2$ & pittsburgh & 5 & 8 & 0.469 & 270.6 & 253.2 & 0.220 & 88.0 \\
H$_2$O & aachen & 5 & 8 & 1.243 & 219.4 & 307.6 & 0.306 & 68.0 \\
H$_2$O & pittsburgh & 5 & 8 & 0.459 & 270.6 & 253.2 & 0.304 & 88.0 \\
\bottomrule
\end{tabular}}
\caption{Calibration of used resources. Means use saved initial-layout qubits and directed calibration entries for couplers in that layout. BeH$_2$ and H$_2$O use five layout qubits forming a four-coupler chain; seven or eight entries are directed calibration records, not distinct undirected edges or a routed-qubit trace. A full post-routing touched-qubit set was not retained, so calibration averages are restricted to the saved initial layout. The Aachen deep cells share session A3. The independently saved Pittsburgh deep-cell snapshots have equal rounded $T_1/T_2$ aggregates but different readout and two-qubit errors. Device-level fields and native basis gates are reported once in Supplementary Table~\ref{tab:qpu-inventory}.}
\label{tab:backends}\label{tab:hardware}
\end{table}

\begin{table}[H]
\centering
\scriptsize
\resizebox{\linewidth}{!}{%
\begin{tabular}{llrrlllrrrrrrl}
\toprule
Backend & Region & Prog. & Qubits+couplers & Family/revision & Basis gates & Status & Pending & 2Q layered & RO error & MCPS & CLOPS & Maintenance & Snapshot\\
\midrule
Phoenix & US & 120 & 338 & Nighthawk r2 & cz,id,rz,sx,x & online & 1 & 3.18e-3 & 7.935e-3 & 109700 & 2000000 & 10/15 & 2026-09-09\\
ibm\_miami & US & 120 & 338 & Nighthawk r1 & not recorded & online & 1 & 9.96e-3 & 1.984e-2 & 250 & 24000 & n/a & 2026-09-09\\
ibm\_boston & US & 156 & 332 & not\_recorded & not recorded & online & 49 & 3.07e-3 & 3.418e-3 & 3800 & 340000 & 10/8 & 2026-09-09\\
ibm\_kingston & US & 156 & 332 & not\_recorded & not recorded & online & 242 & 3.50e-3 & 9.277e-3 & 3800 & 340000 & 10/7 & 2026-09-09\\
Pittsburgh & US & 156 & 332 & Heron r3 & cz,id,rz,sx,x & online & 7 & 3.18e-3 & 3.601e-3 & 3800 & 330000 & 10/13 & 2026-09-09\\
ibm\_fez & US & 156 & 332 & not\_recorded & not recorded & online & 2 & 5.08e-3 & 9.033e-3 & 3900 & 320000 & 10/1 & 2026-09-09\\
ibm\_marrakesh & US & 156 & 332 & not\_recorded & not recorded & online & 9 & 5.00e-3 & 1.27e-2 & 3700 & 300000 & 10/5 & 2026-09-09\\
ibm\_berlin & EU & 120 & 338 & Nighthawk r1 & not recorded & online & 0 & 4.61e-3 & 1.855e-2 & 250 & 25000 & 10/12 & 2026-09-09\\
Aachen & EU & 156 & 332 & Heron r3 & cz,id,rz,sx,x & online & 5 & 3.56e-3 & 5.127e-3 & 3800 & 330000 & 10/6 & 2026-09-09\\
\bottomrule
\end{tabular}}
\caption{Operator-supplied IBM Quantum platform inventory snapshot dated 2026-09-09. Qubits+couplers denotes programmable qubits plus undirected couplers: 332 for 156+176 Heron systems and 338 for 120+218 Nighthawk systems. The operator file supplied 458 for Phoenix; the table corrects this arithmetic inconsistency to 338 because the independently archived 436 directed couplings establish 218 undirected couplers. The supplied value remains preserved in the source CSV for provenance. Basis gates for the three executed backends come from archived backend configurations; other rows were not recorded. The UI reports MCPS rounded to 3800 Hz for Aachen and Pittsburgh, while the archived screening configuration records 3829 Hz. Coupler counts for Nighthawk r1 and the four systems whose family was not recorded are inferred from their qubit-count class. Maintenance dates were displayed as MM/DD without a year and are treated here as 2026. No live service query was made.}
\label{tab:qpu-inventory}
\end{table}

\FloatBarrier
\begin{table}[H]
\centering
\small
\resizebox{\linewidth}{!}{%
\begin{tabular}{llrrrrr}
\toprule
Molecule & Backend & CASCI & Ideal UCC & Noisy Aer & Hardware best & Hardware remeasured \\
\midrule
H$_2$ & aachen & $-1.849662$ & $-1.849662$ & $-1.845254$ & $-1.853022$ & $-1.835895$ \\
H$_2$ & pittsburgh & $-1.849662$ & $-1.849662$ & $-1.845122$ & $-1.853943$ & $-1.824028$ \\
LiH & aachen & $-1.053223$ & $-1.053207$ & $-1.050337$ & $-1.049806$ & $-1.045433$ \\
LiH & pittsburgh & $-1.053223$ & $-1.053207$ & $-1.049261$ & $-1.036556$ & $-1.042760$ \\
HF & aachen & $-1.925584$ & $-1.925584$ & $-1.924747$ & $-1.925759$ & $-1.915814$ \\
HF & pittsburgh & $-1.925584$ & $-1.925584$ & $-1.917907$ & $-1.920604$ & $-1.903872$ \\
O$_2$ & aachen & $-5.362649$ & $-5.362649$ & $-5.319430$ & $-4.896271$ & $-4.774077$ \\
O$_2$ & pittsburgh & $-5.362649$ & $-5.362649$ & $-5.213480$ & $-4.970332$ & $-4.979580$ \\
BeH$_2$ & aachen & $-3.917749$ & $-3.916716$ & $-2.368416$ & $-2.942339$ & $-2.569340$ \\
BeH$_2$ & pittsburgh & $-3.917749$ & $-3.916716$ & $-2.933794$ & $-2.814387$ & $-2.732663$ \\
H$_2$O & aachen & $-6.145814$ & $-6.144190$ & $-4.579314$ & $-4.197818$ & $-4.143084$ \\
H$_2$O & pittsburgh & $-6.145814$ & $-6.144190$ & $-4.468806$ & $-4.201370$ & $-4.077361$ \\
\bottomrule
\end{tabular}%
}
\caption{Active-space energies in hartree. Noisy Aer entries are exact density-matrix expectations at the archived $\theta_{\mathrm{best}}$ vectors, averaged over source vectors; they contain no estimator precision Gaussian and are not trajectory minima. Hardware best denotes the optimizer-selected value, whereas hardware remeasured denotes later resilience-0 fixed-parameter estimator jobs.}
\label{tab:simulation-baselines}
\end{table}

\begin{table*}[!ht]
\centering
\small
\begin{tabular}{llrrr}
\toprule
Backend & Molecule & $n$ & Mean $\Delta E_{\rm ideal}$ & Mean $\Delta E_{\rm noisy}$ \\
\midrule
Aachen & BeH$_2$ & 1 & -0.4843 & -0.3758 \\
Aachen & H$_2$ & 3 & +0.0014 & +0.0014 \\
Aachen & H$_2$O & 1 & +1.4152 & +0.5988 \\
Aachen & HF & 3 & +0.0065 & +0.0065 \\
Aachen & LiH & 4 & +0.0037 & +0.0036 \\
Aachen & O$_2$ & 2 & +0.0217 & +0.0214 \\
Phoenix & LiH & 5 & -0.0018 & -0.0018 \\
Pittsburgh & BeH$_2$ & 1 & +0.6753 & +0.3726 \\
Pittsburgh & H$_2$ & 4 & -0.0407 & -0.0407 \\
Pittsburgh & H$_2$O & 1 & +0.4625 & +0.1346 \\
Pittsburgh & HF & 4 & +0.0003 & +0.0003 \\
Pittsburgh & LiH & 4 & +0.0031 & +0.0030 \\
Pittsburgh & O$_2$ & 5 & +0.0104 & +0.0101 \\
\bottomrule
\end{tabular}
\caption{Offline best--final energy differences at each pair of archived parameter vectors, in mHa. Both columns evaluate the same active-space Hamiltonian. Ideal values use statevectors; noisy values use exact density-matrix expectations with the compact depolarizing models. For the two Heron backends the model uses backend-average archived readout error as a one-qubit depolarizing probability and mean two-qubit gate error as the two-qubit probability. Phoenix uses each source record's archived calibration snapshot and is a separate sensitivity. These surrogates do not model actual readout or the full device noise. No shots are sampled. Entries are means over source records, not uncertainty intervals. Record-level contrasts are rounded to 0.0001 mHa in \texttt{fixed\_theta\_pair\_aer.json}, alongside the model probabilities.}
\label{tab:fixed-theta-pair-aer}
\end{table*}

\begin{figure}[H]
    \centering
    \includegraphics[width=0.92\linewidth]{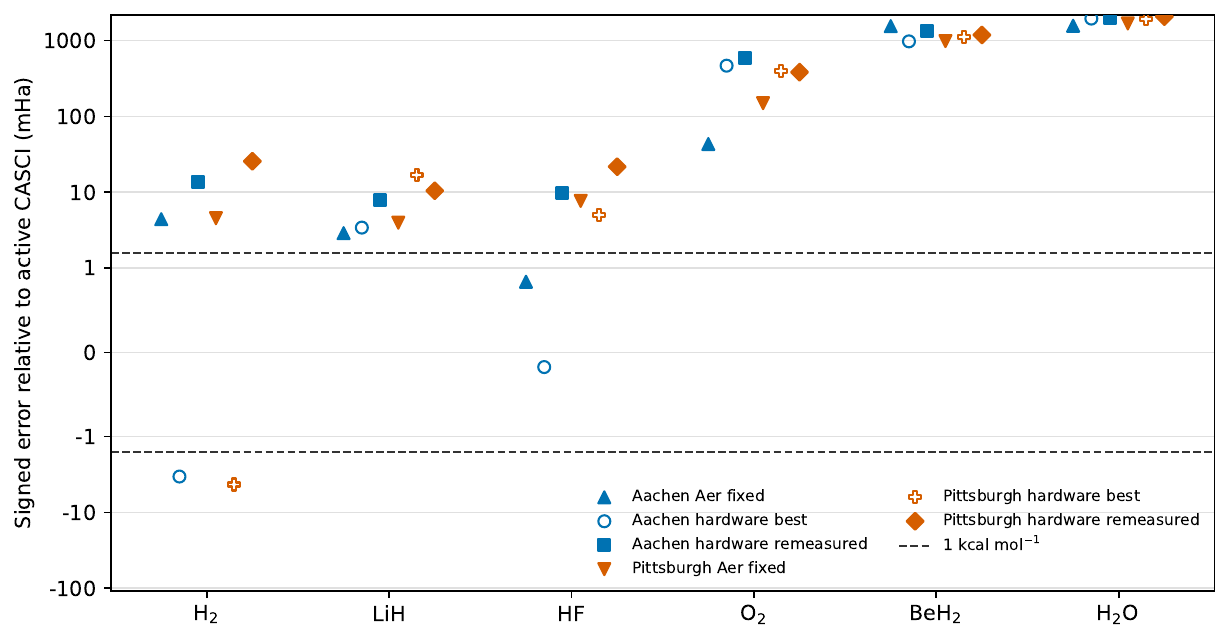}
    \caption{Signed energy errors relative to active-space CASCI on a symlog scale with a 1 mHa linear threshold. Each marker is a mean over source vectors or source records within one molecule/backend cell; no uncertainty bar is shown. Marker shapes distinguish the series and colors distinguish backends; open symbols are optimizer-selected hardware means. Dashed lines mark $\pm1$ kcal mol$^{-1}$. Ordering is descriptive and does not rank devices.}
    \label{fig:simulation-baselines}
\end{figure}
\FloatBarrier
\begin{table}[H]
\centering
\scriptsize
\resizebox{\linewidth}{!}{%
\begin{tabular}{llrrrrr}
\toprule
Dataset & Molecule/backend & $n$ & Signed mean & 95\% CI & One-sided $p$ & TOST $p$ \\
\midrule
Best observed & H$_2$/A & 3 & $-3.36$ & $\pm$ 13.56 & 0.835 & 0.684 \\
Best observed & H$_2$/P & 4 & $-4.28$ & $\pm$ 27.24 & 0.992 & 0.613 \\
Best observed & LiH/A & 4 & +3.42 & $\pm$ 0.99 & 0.995 & 0.995 \\
Best observed & LiH/P & 4 & +16.67 & $\pm$ 2.74 & 1.000 & 1.000 \\
Best observed & HF/A & 3 & $-0.175$ & $\pm$ 5.69 & 0.476 & 0.198 \\
Best observed & HF/P & 4 & +4.98 & $\pm$ 8.76 & 0.860 & 0.847 \\
Best observed & O$_2$/A & 2 & +466.38 & $\pm$ 150.09 & 0.992 & 0.992 \\
Best observed & O$_2$/P & 5 & +392.32 & $\pm$ 96.80 & 1.000 & 1.000 \\
Best observed & BeH$_2$/A & 1 & +975.41 & n/a & n/a & n/a \\
Best observed & BeH$_2$/P & 1 & +1103.36 & n/a & n/a & n/a \\
Best observed & H$_2$O/A & 1 & +1948.00 & n/a & n/a & n/a \\
Best observed & H$_2$O/P & 1 & +1944.44 & n/a & n/a & n/a \\
Remeasured r0 & H$_2$/P & 4 & +25.63 & $\pm$ 7.20 & 0.999 & 0.999 \\
Remeasured r0 & LiH/P & 4 & +10.46 & $\pm$ 1.87 & 1.000 & 1.000 \\
Remeasured r0 & HF/P & 4 & +21.71 & $\pm$ 9.68 & 0.996 & 0.996 \\
Remeasured r0 & O$_2$/P & 5 & +383.07 & $\pm$ 106.03 & 1.000 & 1.000 \\
Remeasured r1 & H$_2$/P & 4 & $-3.33$ & $\pm$ 9.56 & 0.940 & 0.698 \\
Remeasured r1 & LiH/P & 4 & +10.98 & $\pm$ 1.45 & 1.000 & 1.000 \\
Remeasured r1 & HF/P & 4 & +6.46 & $\pm$ 18.30 & 0.866 & 0.770 \\
Remeasured r1 & O$_2$/P & 5 & +435.47 & $\pm$ 107.47 & 1.000 & 1.000 \\
Remeasured r2 & H$_2$/P & 4 & +1.20 & $\pm$ 12.85 & 0.992 & 0.465 \\
Remeasured r2 & LiH/P & 4 & +7.21 & $\pm$ 4.23 & 0.988 & 0.988 \\
Remeasured r2 & HF/P & 4 & +4.90 & $\pm$ 13.36 & 0.826 & 0.756 \\
Remeasured r2 & O$_2$/P & 5 & +150.50 & $\pm$ 127.51 & 0.984 & 0.984 \\
\bottomrule
\end{tabular}}
\caption{Signed errors and chemical-accuracy tests in mHa. The one-sided test uses $H_0:\mu_{|\mathrm{error}|}\geq1.594$ mHa; TOST tests equivalence to $[-1.594,+1.594]$ mHa. No row establishes equivalence. Remeasured and mitigated TOST rows are Pittsburgh-only. The unusually small Aachen HF best-observed mean is retained to three decimals ($-0.175$ mHa) to preserve the calculated mean. A and P denote Aachen and Pittsburgh.}
\label{tab:accuracy-tests}\label{tab:accuracy-tost}
\end{table}

\FloatBarrier

\clearpage
\subsection*{Runtime usage accounting}
Saved Runtime metadata account for \studyQPUSeconds{} QPU s from completed jobs in the primary Heron and stress-test dataset, excluding the timed-out Pittsburgh O$_2$ job. Of this total, \fixedQPUSeconds{} were billed across \allFixedEvaluations{} successful fixed-parameter evaluations and the remainder across its 33 molecule-level VQE records. The post hoc Phoenix work adds \phoenixAllQPU{} billed QPU s from completed jobs: \phoenixSourceQPU{} s for 105 source evaluations, \phoenixFixedQPU{} s for the canonical 100-evaluation fixed acquisition, and \phoenixDuplicateQPU{} s for a completed superseded duplicate retained in the job ledger. The ledger therefore closes under a completed-job convention. Across the primary and Phoenix ledgers, the study total is 9,210 QPU s (2.56 QPU h), including that 12-s superseded duplicate. Phoenix source optimization used one estimator evaluation per job and billed \phoenixSourceRate{} s/job, matching the 1--2-qubit core-job floor rather than the CLOPS-scaled 0.5-s prediction. By contrast, the canonical fixed acquisition batched 50 PUB vectors per job and billed 6 s/job, or \phoenixFixedRate{} s/vector. Post-grouping physical circuit counts are not retained. Queue and wall-clock time are excluded (Supplementary Table~\ref{tab:qpu-usage}).
For the Pittsburgh resilience-level-1 measurements, the pre-submission rates were 3.00, 3.00, 3.00, and 6.00 QPU s per evaluation for H$_2$, LiH, HF, and O$_2$; the measured rates in panel (c) were 2.00, 2.50, 2.00, and 5.30, respectively, so the plans overpredicted usage by factors of 1.13--1.50.
\begin{table}[!htbp]
\centering
\scriptsize
\textbf{(a) Saved billed usage}\par\smallskip
\begin{tabular}{llrr}
\toprule
Backend & Molecule & VQE QPU s & Fixed-parameter QPU s \\
\midrule
Aachen & H$_2$ & 189.0 & 85.0 \\
Aachen & LiH & 270.0 & 199.0 \\
Aachen & HF & 201.0 & 85.0 \\
Aachen & O$_2$ & 288.0 & 324.0 \\
Aachen & BeH$_2$ & 520.0 & 274.0 \\
Aachen & H$_2$O & 720.0 & 366.0 \\
Pittsburgh & H$_2$ & 255.0 & 238.0 \\
Pittsburgh & LiH & 279.0 & 379.0 \\
Pittsburgh & HF & 273.0 & 238.0 \\
Pittsburgh & O$_2$ & 660.0 & 1360.0 \\
Pittsburgh & BeH$_2$ & 420.0 & 300.0 \\
Pittsburgh & H$_2$O & 540.0 & 408.0 \\
Phoenix & LiH (post hoc) & 315.0 & 24.0 \\
Primary total (excludes ibm phoenix) & -- & 4615.0 & 4256.0 \\
\bottomrule
\end{tabular}
\par\medskip
\textbf{(b) Pittsburgh fixed-parameter resilience comparison}\par\smallskip
\resizebox{0.90\linewidth}{!}{%
\begin{tabular}{lrrrrrrrr}
\toprule
Molecule & $n_0$ & $n_{1,2}$ & r0 error & r1 error & r2 error & $\Delta$(1--0) & $\Delta$(2--0) & r2 $\sigma$ \\
\midrule
H$_2$ & 40 & 12 & 25.63 & $-3.33$ & 1.20 & $-28.96$ & $-24.43$ & 12.30 \\
LiH & 40 & 12 & 10.46 & 10.98 & 7.21 & 0.52 & $-3.26$ & 4.14 \\
HF & 40 & 12 & 21.71 & 6.46 & 4.90 & $-15.25$ & $-16.81$ & 9.25 \\
O$_2$ & 50 & 15 & 383.07 & 435.47 & 150.50 & 52.40 & $-232.57$ & 97.69 \\
\bottomrule
\end{tabular}}
\par\medskip
\textbf{(c) Mitigation cost-benefit}\par\smallskip
\resizebox{0.90\linewidth}{!}{%
\begin{tabular}{lrrrr}
\toprule
Molecule & Resilience & QPU s/eval & Absolute error (mHa) & Improvement vs r0 (mHa) \\
\midrule
H$_2$ & 0 & 0.72 & 25.63 & +0.00 \\
H$_2$ & 1 & 2.00 & 3.33 & +22.31 \\
H$_2$ & 2 & 3.12 & 1.20 & +24.43 \\
LiH & 0 & 1.24 & 10.46 & +0.00 \\
LiH & 1 & 2.50 & 10.98 & $-0.52$ \\
LiH & 2 & 5.00 & 7.21 & +3.26 \\
HF & 0 & 0.72 & 21.71 & +0.00 \\
HF & 1 & 2.00 & 6.46 & +15.25 \\
HF & 2 & 3.12 & 4.90 & +16.81 \\
O$_2$ & 0 & 3.92 & 383.07 & +0.00 \\
O$_2$ & 1 & 5.30 & 435.47 & $-52.40$ \\
O$_2$ & 2 & 13.90 & 150.50 & +232.57 \\
\bottomrule
\end{tabular}}
\caption{Runtime usage and mitigation sensitivity. Panel (a) reports saved billed QPU seconds and excludes queue and wall-clock time. The Phoenix fixed-parameter entry of 24.0 QPU s comprises the canonical 12.0-s acquisition (100 evaluations) and a 12.0-s superseded duplicate; only the canonical acquisition enters quantitative analysis, while $315+24=339$ QPU s matches the separately scoped Phoenix usage in Methods. Panel (b) gives signed CASCI errors in mHa for $\theta_{\rm best}$. In panel (c), resilience-0 counts are 160 evaluations for each shallow molecule and 200 for O$_2$, covering both vectors and both epochs. The displayed $n_{1,2}=12,12,12,15$ counts H$_2$, LiH, HF, and O$_2$ $\theta_{\rm best}$ evaluations at each mitigated level; each level contributes $2n_{1,2}$ evaluations after including $\theta_{\rm final}$. These counts and measured rates reproduce the panel-(a) Pittsburgh fixed totals 238.0, 379.0, 238.0, and 1360.0 QPU s. Positive improvement denotes reduced absolute error.}
\label{tab:qpu-usage}\label{tab:resilience-comparison}\label{tab:mitigation-cost-benefit}
\end{table}

\FloatBarrier

\clearpage
\subsection*{Optimization plateau fluctuations}
For each successful run, the sample standard deviation is computed separately over the final 30\% of objective evaluations and is never pooled across runs. The table reports the median per-run value and min--max range; no range is reported for single-run groups. Coefficient-norm quantities remain proxies because the required allocation and covariance metadata are unavailable.
\begin{table}[H]
\centering
\tiny
\textbf{(a) Source-trajectory plateau scatter}\par\smallskip
\resizebox{\linewidth}{!}{%
\begin{tabular}{llrrrrrr}
\toprule
Molecule & Backend & Runs & Evals/run & Plateau $\sigma$ & $L_2/\sqrt S$ & Unweighted $L_1/\sqrt S$ & Points/run \\
\midrule
H$_2$ & Aachen & 3 & 22,21,20 & 6.711 [6.029, 6.787] & 25.216 & 30.238 & 6 \\
H$_2$ & Pittsburgh & 4 & 23,18,23,21 & 9.843 [7.364, 12.910] & 25.216 & 30.238 & 5-6 \\
LiH & Aachen & 4 & 21,22,21,27 & 1.155 [0.878, 1.503] & 8.284 & 17.548 & 6-8 \\
LiH & Pittsburgh & 4 & 23,20,26,24 & 5.847 [1.832, 10.548] & 8.284 & 17.548 & 6-7 \\
HF & Aachen & 3 & 22,20,25 & 2.347 [1.811, 2.565] & 20.623 & 21.587 & 6-7 \\
HF & Pittsburgh & 4 & 23,25,23,20 & 5.777 [3.864, 6.082] & 20.623 & 21.587 & 6-7 \\
O$_2$ & Aachen & 2 & 23,25 & 21.382 [15.720, 27.043] & 25.066 & 75.888 & 6-7 \\
O$_2$ & Pittsburgh & 5 & 22,22,25,22,19 & 134.037 [63.688, 142.217] & 25.066 & 75.888 & 5-7 \\
BeH$_2$ & Aachen & 1 & 40 & 30.328 n/a & 24.166 & 111.803 & 12 \\
BeH$_2$ & Pittsburgh & 1 & 30 & 24.782 n/a & 24.166 & 111.803 & 9 \\
H$_2$O & Aachen & 1 & 40 & 50.430 n/a & 40.903 & 202.213 & 12 \\
H$_2$O & Pittsburgh & 1 & 30 & 61.674 n/a & 40.903 & 202.213 & 9 \\
\bottomrule
\end{tabular}}
\par\medskip
\textbf{(b) Fixed-parameter scatter and coefficient proxies}\par\smallskip
\begin{tabular}{llrrrr}
\toprule
Backend & Molecule & $\tilde\sigma_{bf}$ & $L_2/\sqrt S$ & Ratio & Unweighted $L_1/\sqrt S$ \\
\midrule
A & H$_2$ & 6.73 & 25.22 & 0.27 & 30.24 \\
A & LiH & 1.71 & 8.28 & 0.21 & 17.55 \\
A & HF & 3.31 & 20.62 & 0.16 & 21.59 \\
A & O$_2$ & 20.36 & 25.07 & 0.81 & 75.89 \\
A & BeH$_2$ & 19.91 & 24.17 & 0.82 & 111.80 \\
A & H$_2$O & 50.18 & 40.90 & 1.23 & 202.21 \\
P & H$_2$ & 11.67 & 25.22 & 0.46 & 30.24 \\
P & LiH & 2.14 & 8.28 & 0.26 & 17.55 \\
P & HF & 5.82 & 20.62 & 0.28 & 21.59 \\
P & O$_2$ & 32.85 & 25.07 & 1.31 & 75.89 \\
P & BeH$_2$ & 21.70 & 24.17 & 0.90 & 111.80 \\
P & H$_2$O & 44.48 & 40.90 & 1.09 & 202.21 \\
pooled & H$_2$ & 10.74 & 25.22 & 0.43 & 30.24 \\
pooled & LiH & 2.05 & 8.28 & 0.25 & 17.55 \\
pooled & HF & 4.14 & 20.62 & 0.20 & 21.59 \\
pooled & O$_2$ & 30.05 & 25.07 & 1.20 & 75.89 \\
pooled & BeH$_2$ & 20.81 & 24.17 & 0.86 & 111.80 \\
pooled & H$_2$O & 50.18 & 40.90 & 1.23 & 202.21 \\
\bottomrule
\end{tabular}
\caption{Noise-scale diagnostics in mHa at 1024 shots. Panel (a) reports the median per-run SD over each trajectory's final 30\% with its min--max range; evaluation counts are ordered chronologically by session key within each molecule/backend cell. Panel (b) reports $\tilde\sigma_{bf}$, the median of within-vector SDs across both best and final vectors from the two primary Heron backends. A and P denote Aachen and Pittsburgh; pooled combines both. The $L_1$ columns are unweighted coefficient norms; Supplementary Table~\ref{tab:selection-decomposition} instead uses the Pauli-weighted norm $\sum|c_k|w_k$. Coefficient norms remain proxies because shot-allocation and commuting-group covariance are unavailable; panel (b) supplies the descriptive ratios and precision-planning input.}
\label{tab:hardware-plateau-sigma}\label{tab:remeasurement-noise}\label{tab:noise-proxy-ratios}
\end{table}

\FloatBarrier
\begingroup
\tiny
\setlength{\tabcolsep}{2pt}
\begin{longtable}{lllllrrrr}
\toprule
Molecule & Backend & Session & Run ID & Runtime & Best & Final & Drift & Evals/maxiter \\
\midrule
\endfirsthead
\toprule
Molecule & Backend & Session & Run ID & Runtime & Best & Final & Drift & Evals/maxiter \\
\midrule
\endhead
LiH & aachen & A1 & 2026.07.30-21.17.21 & 0.40.1 & $-1.049464$ & $-1.046120$ & 3.34 & 21/40 \\
H$_2$ & aachen & A2 & 2026-08-01-09.02.35 & 0.40.1 & $-1.847906$ & $-1.832080$ & 15.83 & 22/40 \\
HF & aachen & A2 & 2026-08-01-09.02.35 & 0.40.1 & $-1.923601$ & $-1.919853$ & 3.75 & 22/40 \\
LiH & aachen & A2 & 2026-08-01-09.02.35 & 0.40.1 & $-1.050219$ & $-1.046043$ & 4.18 & 22/40 \\
BeH$_2$ & aachen & A3 & 2026-08-01-12.11.09 & 0.40.1 & $-2.942339$ & $-2.863384$ & 78.96 & 40/40 \\
H$_2$O & aachen & A3 & 2026-08-01-12.11.09 & 0.40.1 & $-4.197818$ & $-4.007231$ & 190.59 & 40/40 \\
O$_2$ & aachen & A3 & 2026-08-01-12.11.09 & 0.40.1 & $-4.884458$ & $-4.875001$ & 9.46 & 23/40 \\
H$_2$ & aachen & A4 & 2026-08-01-16.36.57 & 0.40.1 & $-1.852392$ & $-1.841409$ & 10.98 & 21/40 \\
HF & aachen & A4 & 2026-08-01-16.36.57 & 0.40.1 & $-1.925515$ & $-1.920100$ & 5.41 & 20/40 \\
LiH & aachen & A4 & 2026-08-01-16.36.57 & 0.40.1 & $-1.049108$ & $-1.044650$ & 4.46 & 21/40 \\
O$_2$ & aachen & A4 & 2026-08-01-16.36.57 & 0.40.1 & $-4.908083$ & $-4.873294$ & 34.79 & 25/40 \\
H$_2$ & aachen & A5 & 2026-08-01-18.57.40 & 0.40.1 & $-1.858767$ & $-1.839636$ & 19.13 & 20/40 \\
HF & aachen & A5 & 2026-08-01-18.57.40 & 0.40.1 & $-1.928161$ & $-1.921767$ & 6.39 & 25/40 \\
LiH & aachen & A5 & 2026-08-01-18.57.40 & 0.40.1 & $-1.050433$ & $-1.047592$ & 2.84 & 27/40 \\
H$_2$ & pittsburgh & P1 & 2026.07.30-22.13.56 & 0.40.1 & $-1.859460$ & $-1.824027$ & 35.43 & 23/40 \\
LiH & pittsburgh & P2 & 2026.07.31-03.08.12 & 0.40.1 & $-1.036633$ & $-1.033085$ & 3.55 & 23/40 \\
HF & pittsburgh & P3 & 2026.07.31-18.20.56 & 0.40.1 & $-1.914220$ & $-1.891541$ & 22.68 & 23/40 \\
H$_2$ & pittsburgh & P4 & 2026-08-01-12.09.37 & 0.40.1 & $-1.828758$ & $-1.828758$ & 0.00 & 18/40 \\
HF & pittsburgh & P4 & 2026-08-01-12.09.37 & 0.40.1 & $-1.925842$ & $-1.919096$ & 6.75 & 25/40 \\
LiH & pittsburgh & P4 & 2026-08-01-12.09.37 & 0.40.1 & $-1.036172$ & $-1.034790$ & 1.38 & 20/40 \\
H$_2$ & pittsburgh & P5 & 2026-08-01-17.54.22 & 0.40.1 & $-1.860528$ & $-1.833957$ & 26.57 & 23/40 \\
HF & pittsburgh & P5 & 2026-08-01-17.54.22 & 0.40.1 & $-1.917852$ & $-1.909942$ & 7.91 & 23/40 \\
LiH & pittsburgh & P5 & 2026-08-01-17.54.22 & 0.40.1 & $-1.038796$ & $-1.033563$ & 5.23 & 26/40 \\
O$_2$ & pittsburgh & P6 & 2026-08-01-18.56.23 & 0.40.1 & $-4.879641$ & $-4.838620$ & 41.02 & 22/40 \\
O$_2$ & pittsburgh & P7 & 2026-08-02-07.22.35 & 0.40.1 & $-4.982996$ & $-4.917501$ & 65.50 & 22/40 \\
O$_2$ & pittsburgh & P8 & 2026-08-02-16.04.58 & 0.40.1 & $-5.017644$ & $-4.772132$ & 245.51 & 25/40 \\
H$_2$ & pittsburgh & P9 & 2026-08-02-18.02.42 & 0.40.1 & $-1.867025$ & $-1.856752$ & 10.27 & 21/40 \\
HF & pittsburgh & P9 & 2026-08-02-18.02.42 & 0.40.1 & $-1.924502$ & $-1.914573$ & 9.93 & 20/40 \\
LiH & pittsburgh & P9 & 2026-08-02-18.02.42 & 0.40.1 & $-1.034622$ & $-1.029864$ & 4.76 & 24/40 \\
O$_2$ & pittsburgh & P9 & 2026-08-02-18.02.42 & 0.40.1 & $-4.904397$ & $-4.808553$ & 95.84 & 22/40 \\
O$_2$ & pittsburgh & P10 & 2026-08-15-22.13.13 & 0.40.1 & $-5.066983$ & $-4.984992$ & 81.99 & 19/40 \\
H$_2$O & pittsburgh & P11 & 2026-09-01-17.16.37 & 0.49.0 & $-4.201370$ & $-4.147017$ & 54.35 & 30/30 \\
BeH$_2$ & pittsburgh & P12 & 2026-09-02-06.17.58 & 0.49.0 & $-2.814387$ & $-2.745518$ & 68.87 & 30/30 \\
\bottomrule
\caption{Per-record optimizer outcomes. Best and final energies are in hartree; drift is final minus best in mHa. Session keys identify records acquired in the same hardware invocation. Runtime versions are reconstructed from saved transpilation metadata.}\label{tab:run-details}\\
\end{longtable}
\endgroup

\begingroup
\tiny
\setlength{\tabcolsep}{2pt}
\begin{longtable}{lllrrrrrr}
\toprule
Molecule & Backend & Session & Source & Fixed & Shift & $m$ & $s_b$ & $b_i$ \\
\midrule
\endfirsthead
\toprule
Molecule & Backend & Session & Source & Fixed & Shift & $m$ & $s_b$ & $b_i$ \\
\midrule
\endhead
LiH & aachen & A1 & $-1.049464$ & $-1.046399$ & +3.06 & 21 & 1.74 & 3.29 \\
H$_2$ & aachen & A2 & $-1.847906$ & $-1.834653$ & +13.25 & 22 & 3.10 & 5.92 \\
HF & aachen & A2 & $-1.923601$ & $-1.915424$ & +8.18 & 22 & 4.53 & 8.65 \\
LiH & aachen & A2 & $-1.050219$ & $-1.046259$ & +3.96 & 22 & 1.70 & 3.25 \\
BeH$_2$ & aachen & A3 & $-2.942339$ & $-2.569340$ & +373.00 & 40 & 21.68 & 46.75 \\
H$_2$O & aachen & A3 & $-4.197818$ & $-4.143084$ & +54.73 & 40 & 54.56 & 117.65 \\
O$_2$ & aachen & A3 & $-4.884458$ & $-4.782907$ & +101.55 & 23 & 13.44 & 25.93 \\
H$_2$ & aachen & A4 & $-1.852392$ & $-1.832775$ & +19.62 & 21 & 7.36 & 13.91 \\
HF & aachen & A4 & $-1.925515$ & $-1.916270$ & +9.25 & 20 & 3.04 & 5.68 \\
LiH & aachen & A4 & $-1.049108$ & $-1.042586$ & +6.52 & 21 & 2.61 & 4.93 \\
O$_2$ & aachen & A4 & $-4.908083$ & $-4.765246$ & +142.84 & 25 & 22.43 & 44.06 \\
H$_2$ & aachen & A5 & $-1.858767$ & $-1.840257$ & +18.51 & 20 & 6.62 & 12.36 \\
HF & aachen & A5 & $-1.928161$ & $-1.915748$ & +12.41 & 25 & 3.71 & 7.29 \\
LiH & aachen & A5 & $-1.050433$ & $-1.046489$ & +3.94 & 27 & 1.68 & 3.36 \\
H$_2$ & pittsburgh & P1 & $-1.859460$ & $-1.820552$ & +38.91 & 23 & 11.41 & 22.01 \\
LiH & pittsburgh & P2 & $-1.036633$ & $-1.042400$ & $-5.77$ & 23 & 2.03 & 3.91 \\
HF & pittsburgh & P3 & $-1.914220$ & $-1.895054$ & +19.17 & 23 & 6.34 & 12.22 \\
H$_2$ & pittsburgh & P4 & $-1.828758$ & $-1.820347$ & +8.41 & 18 & 12.39 & 22.57 \\
HF & pittsburgh & P4 & $-1.925842$ & $-1.906913$ & +18.93 & 25 & 5.90 & 11.59 \\
LiH & pittsburgh & P4 & $-1.036172$ & $-1.041250$ & $-5.08$ & 20 & 2.08 & 3.88 \\
H$_2$ & pittsburgh & P5 & $-1.860528$ & $-1.829867$ & +30.66 & 23 & 10.83 & 20.88 \\
HF & pittsburgh & P5 & $-1.917852$ & $-1.908681$ & +9.17 & 23 & 6.55 & 12.63 \\
LiH & pittsburgh & P5 & $-1.038796$ & $-1.043720$ & $-4.92$ & 26 & 2.16 & 4.29 \\
O$_2$ & pittsburgh & P6 & $-4.879641$ & $-5.002543$ & $-122.90$ & 22 & 28.19 & 53.83 \\
O$_2$ & pittsburgh & P7 & $-4.982996$ & $-4.979291$ & +3.71 & 22 & 37.97 & 72.51 \\
O$_2$ & pittsburgh & P8 & $-5.017644$ & $-5.024261$ & $-6.62$ & 25 & 42.84 & 84.15 \\
H$_2$ & pittsburgh & P9 & $-1.867025$ & $-1.825344$ & +41.68 & 21 & 12.90 & 24.38 \\
HF & pittsburgh & P9 & $-1.924502$ & $-1.904841$ & +19.66 & 20 & 5.73 & 10.71 \\
LiH & pittsburgh & P9 & $-1.034622$ & $-1.043670$ & $-9.05$ & 24 & 2.32 & 4.51 \\
O$_2$ & pittsburgh & P9 & $-4.904397$ & $-4.835501$ & +68.90 & 22 & 27.25 & 52.04 \\
O$_2$ & pittsburgh & P10 & $-5.066983$ & $-5.056303$ & +10.68 & 19 & 31.91 & 58.90 \\
H$_2$O & pittsburgh & P11 & $-4.201370$ & $-4.077361$ & +124.01 & 30 & 58.18 & 118.71 \\
BeH$_2$ & pittsburgh & P12 & $-2.814387$ & $-2.732663$ & +81.72 & 30 & 19.95 & 40.69 \\
\bottomrule
\caption{Run-level resilience-0 $\theta_\mathrm{best}$ remeasurement on both backends. Fixed is the later-epoch mean and shift is fixed minus source. For each run, $m$ is the source-trajectory length, $s_b$ is the sample SD of repeated best-point measurements, and $b_i=a_{m_i}s_{b,i}$; hence every cell prediction $b=\operatorname{mean}(b_i)$ is reproducible. Energies are in hartree and other quantities in mHa. Session keys match Supplementary Table~\ref{tab:run-details}.}\label{tab:paired-remeasurement}\label{tab:aachen-paired}\\
\end{longtable}
\endgroup

\clearpage
\subsection*{Remeasurement noise and selection diagnostics}
Enumerated session-level wild-cluster Rademacher intervals replace unclustered intervals for primary uncertainty reporting; the core and extended results are given in Supplementary Table~\ref{tab:cluster-inference}.
Supplementary Tables~\ref{tab:remeasurement-noise}--\ref{tab:noise-floor-requirements} report the measured within-vector scatter, acquisition lag and calibration changes, order-statistic sensitivity, and precision requirements.
\begin{table}[H]
\centering
\tiny
\resizebox{\linewidth}{!}{%
\begin{tabular}{llrrrrrrrrrrrrr}
\toprule
& & \multicolumn{3}{c}{Acquisition} & \multicolumn{4}{c}{Noise} & \multicolumn{4}{c}{Selection} & \\
Backend & Mol. & Lag (d) & $\Delta$RO & $\Delta$2q & $\tilde\sigma_p$ & $\tilde\sigma_b$ & $R$ & rel. SE & $b$ & $d_{sel}$ & $d_{sel}/b$ & Env. & Sign \\
\midrule
A & H$_2$ & 36.8 & +0.00818 & n/a & 6.71 & 6.62 & 1.01 & 0.39 & 10.7 & 13.5 & 1.26 & 15.8 & $+$ \\
A & LiH & 37.1 & +0.00494 & +0.00007 & 1.16 & 1.72 & 0.67 & 0.39 & 3.7 & 3.8 & 1.02 & 7.2 & $+$ \\
A & HF & 36.9 & +0.00818 & n/a & 2.35 & 3.71 & 0.63 & 0.39 & 7.2 & 6.4 & 0.89 & 11.3 & $+$ \\
A & O$_2$ & 37.1 & +0.00737 & $-0.00005$ & 21.38 & 17.94 & 1.19 & 0.39 & 35.0 & 25.1 & 0.72 & 79.5 & $+$ \\
A & BeH$_2$ & 38.2 & +0.00620 & $-0.00206$ & 30.33 & 21.68 & 1.40 & 0.30 & 46.8 & 71.2 & 1.52 & 101.4 & $+$ \\
A & H$_2$O & 38.4 & +0.00620 & $-0.00177$ & 50.43 & 54.56 & 0.92 & 0.30 & 117.6 & 210.3 & 1.79 & 195.8 & $+$ \\
P & H$_2$ & 29.9 & +0.00397 & n/a & 11.06 & 11.41 & 0.97 & 0.39 & 22.4 & 23.7 & 1.06 & 7.7 & $+$ \\
P & LiH & 29.9 & +0.00299 & $-0.00050$ & 5.85 & 2.12 & 2.76 & 0.39 & 4.1 & 3.5 & 0.85 & 4.3 & $-$ \\
P & HF & 29.9 & +0.00397 & n/a & 5.78 & 6.12 & 0.94 & 0.39 & 11.8 & 9.0 & 0.76 & 5.5 & $+$ \\
P & O$_2$ & 28.8 & +0.00314 & $-0.00079$ & 134.04 & 31.91 & 4.20 & 0.42 & 64.3 & 99.5 & 1.55 & 33.9 & $-$ \\
P & BeH$_2$ & 5.6 & $-0.00034$ & +0.00036 & 24.78 & 19.95 & 1.24 & 0.33 & 40.7 & 62.2 & 1.53 & 5.6 & $+$ \\
P & H$_2$O & 6.1 & $-0.00024$ & $-0.00079$ & 61.67 & 58.18 & 1.06 & 0.33 & 118.7 & 45.9 & 0.39 & 7.7 & $+$ \\
\bottomrule
\end{tabular}}
\caption{Per-cell acquisition, noise, and selection diagnostics. Lag is the median source-to-remeasurement interval within each cell. Errors and noise scales are in mHa; $\Delta$RO and $\Delta$2q are probability fractions. $\tilde\sigma_p$, $\tilde\sigma_b$, and the plateau-size input $n_p$ all exclude the Pittsburgh H$_2$ zero-source-gap diagnostic; Supplementary Table~\ref{tab:hardware-plateau-sigma} retains that control and therefore reports 9.843 mHa for Pittsburgh H$_2$. $R=\tilde\sigma_p/\tilde\sigma_b$. Relative SE is $\{[2(n_p-1)]^{-1}+[2(k-1)]^{-1}\}^{1/2}$ using the minimum included plateau size $n_p$ and fixed sample size $k$. Sign denotes the raw best-point remeasurement shift. $\mathrm{Env}=2|\Delta\epsilon_{\rm RO}|\sum_{k\ne I}|c_k|w_k$, where $w_k$ is Pauli weight and the identity is excluded. The Pauli-weighted norms $\sum|c_k|w_k$ are H$_2$ 967.6, LiH 725.9, HF 690.8, O$_2$ 5395.6, BeH$_2$ 8175.4, H$_2$O 15785.8 mHa; they equal the unweighted $L_1$ norm for one-qubit cells and are 1.3--2.4 times larger otherwise. Eight of twelve raw best-point shift magnitudes and six of twelve $d_{\rm sel}$ magnitudes exceed Env. A/P denote Aachen/Pittsburgh.}
\label{tab:selection-decomposition}\label{tab:remeasurement-lag}\label{tab:plateau-noise-ratio}\label{tab:readout-drift-envelope}\label{tab:shift-sign-concordance}
\end{table}

\FloatBarrier
Using source-plateau scatter in the order-statistic scale gives a mean cell ratio of \plateauRatioMean{} (between-cell SD \plateauRatioSD{}; cell-$t$ interval $[\plateauRatioCILow{},\plateauRatioCIHigh{}]$) across the 12 cells. This sensitivity changes the noise epoch, not the full-trajectory competitor count; the interval describes between-cell variation rather than the primary session-clustered uncertainty.
\begin{table}[H]
\centering
\scriptsize
\textbf{(a) Order-statistic sensitivity}\par\smallskip
\resizebox{\linewidth}{!}{%
\begin{tabular}{llrrrrr}
\toprule
Backend & Molecule & $d_{\rm sel}$ & $b_{\rm fixed}$ & $d_{\rm sel}/b_f$ & $b_{\rm plateau}$ & $d_{\rm sel}/b_p$ \\
\midrule
Aachen & H$_2$ & 13.52 & 10.73 & 1.26 & 12.29 & 1.10 \\
Aachen & LiH & 3.79 & 3.71 & 1.02 & 2.25 & 1.69 \\
Aachen & HF & 6.44 & 7.21 & 0.89 & 4.30 & 1.50 \\
Aachen & O$_2$ & 25.13 & 35.00 & 0.72 & 41.72 & 0.60 \\
Aachen & BeH$_2$ & 71.16 & 46.75 & 1.52 & 65.40 & 1.09 \\
Aachen & H$_2$O & 210.27 & 117.65 & 1.79 & 108.74 & 1.93 \\
Pittsburgh & H$_2$ & 23.67 & 22.43 & 1.06 & 20.05 & 1.18 \\
Pittsburgh & LiH & 3.52 & 4.15 & 0.85 & 11.73 & 0.30 \\
Pittsburgh & HF & 9.02 & 11.79 & 0.76 & 10.36 & 0.87 \\
Pittsburgh & O$_2$ & 99.50 & 64.29 & 1.55 & 221.92 & 0.45 \\
Pittsburgh & BeH$_2$ & 62.17 & 40.69 & 1.53 & 50.56 & 1.23 \\
Pittsburgh & H$_2$O & 45.88 & 118.71 & 0.39 & 125.83 & 0.36 \\
\bottomrule
\end{tabular}}
\par\medskip
\textbf{(b) Precision-planning counts}\par\smallskip
\begin{tabular}{lrr}
\toprule
Molecule & Pooled $\tilde\sigma_{bf}$ (mHa) & Evaluations \\
\midrule
H$_2$ & 10.74 & 182 \\
LiH & 2.05 & 7 \\
HF & 4.14 & 28 \\
O$_2$ & 30.05 & 1422 \\
BeH$_2$ & 20.81 & 682 \\
H$_2$O & 50.18 & 3964 \\
\bottomrule
\end{tabular}
\par\medskip
\textbf{(c) Independence-reference sensitivity for replicated cells}\par\smallskip
\resizebox{\linewidth}{!}{%
\begin{tabular}{llrrrrr}
\toprule
Backend & Molecule & Record-level $d_{\rm sel}$ SD & Independence reference & Ratio & One-sided $p$ & Holm $p$ \\
\midrule
Aachen & H$_2$ & 3.84 & 6.24 & 0.62 & 0.0321 & 0.0963 \\
Aachen & LiH & 0.66 & 2.12 & 0.31 & 0.0186 & 0.0746 \\
Aachen & HF & 0.88 & 4.12 & 0.21 & 0.0568 & 0.1137 \\
Aachen & O$_2$ & 8.75 & 19.65 & 0.45 & 0.1608 & 0.1608 \\
Pittsburgh & H$_2$ & 9.09 & 12.83 & 0.71 & 0.0428 & 0.0867 \\
Pittsburgh & LiH & 1.75 & 2.35 & 0.74 & 0.0289 & 0.0867 \\
Pittsburgh & HF & 4.62 & 6.71 & 0.69 & 0.0373 & 0.0867 \\
Pittsburgh & O$_2$ & 81.34 & 36.84 & 2.21 & 0.0019 & 0.0076 \\
\bottomrule
\end{tabular}}
\caption{Selection-model sensitivity and precision planning. Panel (a) compares $d_{\rm sel}$ with predictions based on later fixed-parameter scatter and source plateau scatter. Here $b_{\rm plateau}=\operatorname{mean}_i(a_{m_i}s_{p,i})$, using each record's full trajectory length $m_i$ and source-plateau scatter $s_{p,i}$. Panel (b) takes the primary Heron pooled $\tilde\sigma_{bf}$ from Supplementary Table~\ref{tab:remeasurement-noise}. Counts are $\lfloor(\tilde\sigma_{bf}/0.797)^2\rfloor+1$ from unrounded values, so a count inferred from the displayed SD can differ by one. Panel (c)'s record-level $d_{\rm sel}$ SD is reproducible from Supplementary Table~\ref{tab:theta-final-control} as $\mathrm{SD}=h\sqrt{n}/t_{n-1}$, where $h$ is its 95\% half-width; for Aachen H$_2$, $9.54\sqrt{3}/4.3027=3.84$ mHa. It is not the cell-mean shift in main-text Table~\ref{tab:fixed-validation}. The reference $s_b\sqrt{1+2/k}$ treats the selected source-best value and parameter-dependent means as fixed; random terms are the single source-final evaluation with variance $s_b^2$ and two independent $k$-evaluation fixed means with variance $s_b^2/k$ each. A four-term independence model instead gives $s_b\sqrt{2}\sqrt{1+1/k}=1.483s_b$ at $k=10$, 1.354 times the displayed reference, so every ratio would be divided by 1.354. Its t tests are assumption-dependent sensitivity analyses; only Pittsburgh O$_2$ survives within-backend Holm adjustment.}
\label{tab:dsel-robustness}\label{tab:noise-floor-requirements}
\end{table}

\begin{table}[!ht]
\centering
\small
\begin{tabular}{llrrrr}
\toprule
Backend & Molecule & $n$ & Mean $m$ & $d_{\rm sel}/b_{\rm full}$ & $d_{\rm sel}/b_{\rm plateau\ count}$ \\
\midrule
Aachen & BeH$_2$ & 1 & 40.0 & 1.522 & 2.007 \\
Aachen & H$_2$ & 3 & 21.0 & 1.260 & 1.853 \\
Aachen & H$_2$O & 1 & 40.0 & 1.787 & 2.357 \\
Aachen & HF & 3 & 22.3 & 0.893 & 1.308 \\
Aachen & LiH & 4 & 22.8 & 1.023 & 1.491 \\
Aachen & O$_2$ & 2 & 24.0 & 0.718 & 1.050 \\
Pittsburgh & BeH$_2$ & 1 & 30.0 & 1.528 & 2.086 \\
Pittsburgh & H$_2$ & 3 & 22.3 & 1.055 & 1.576 \\
Pittsburgh & H$_2$O & 1 & 30.0 & 0.386 & 0.528 \\
Pittsburgh & HF & 4 & 22.8 & 0.765 & 1.130 \\
Pittsburgh & LiH & 4 & 23.2 & 0.849 & 1.239 \\
Pittsburgh & O$_2$ & 5 & 22.0 & 1.548 & 2.305 \\
\bottomrule
\end{tabular}
\caption{Competitor-count sensitivity at fixed later-epoch best-vector scatter. The primary full-trajectory count is the saved evaluation count $m$, because every evaluated point could set the reported minimum. The secondary count is the number of final-plateau evaluations, $m-\lceil0.70m\rceil$. Both use Blom's $a_m$ separately for each source run, then average $a_m s_{b,i}$ within a cell before taking $d_{\rm sel}/b$. This changes the competitor count only; it is distinct from substituting source-plateau scatter for later fixed-job scatter. Neither count choice was preregistered. Singleton stress cells have no between-run uncertainty.}
\label{tab:competitor-count}
\end{table}

\begin{table}[!ht]
\centering
\small
\begin{tabular}{lrrr}
\toprule
Comparison & Records & Standardized slope & Permutation $p$ \\
\midrule
dsel vs lag & 28 & -0.026 & 0.9079 \\
dsel vs readout change & 22 & +0.195 & 0.4449 \\
dsel vs two-qubit error change & 13 & +0.030 & 0.9012 \\
source drift vs theta distance & 22 & -0.279 & 0.2645 \\
\bottomrule
\end{tabular}
\caption{Exploratory within-cell standardized linear associations in the replicated source set. For each comparison, cells with no within-cell predictor or outcome variation are omitted. Two-sided Monte Carlo $p$ values permute the predictor within backend--molecule cells; they are not session-cluster intervals, are not adjusted for multiple comparisons, and cannot identify causation. Lag uses the saved source-run identifier and median first-campaign fixed-job creation time. Readout and two-qubit error changes subtract saved source calibration averages from the first fixed campaign averages. The two-qubit comparison excludes cells without used-edge error fields. Source theta distances come from the saved-parameter ideal-gap audit.}
\label{tab:exploratory-associations}
\end{table}

\clearpage
\begin{table}[H]
\centering
\tiny
\textbf{(a) Final-point and drift-differenced controls}\par\smallskip
\resizebox{\linewidth}{!}{%
\begin{tabular}{llrrrrrrrrrr}
\toprule
& & \multicolumn{5}{c}{$\theta_{\rm final}$ shift} & \multicolumn{5}{c}{$d_{\rm sel}$} \\
Molecule & Backend & $n_f$ & Mean & 95\% half-width & $p$ & Holm $p$ & $n_d$ & Mean & 95\% half-width & $p$ & Holm $p$ \\
\midrule
H$_2$ & A & 3 & +3.61 & 14.03 & 0.3838 & 0.7676 & 3 & +13.52 & 9.54 & 0.0259 & 0.0518 \\
LiH & A & 4 & +0.58 & 2.45 & 0.5055 & 0.7676 & 4 & +3.79 & 1.04 & 0.0014 & 0.0056 \\
HF & A & 3 & +3.51 & 3.41 & 0.0475 & 0.1899 & 3 & +6.44 & 2.18 & 0.0062 & 0.0185 \\
O$_2$ & A & 2 & +97.07 & 183.70 & 0.0941 & 0.2824 & 2 & +25.13 & 78.59 & 0.1537 & 0.1537 \\
BeH$_2$ & A & 1 & +301.84 & n/a & n/a & n/a & 1 & +71.16 & n/a & n/a & n/a \\
H$_2$O & A & 1 & $-155.53$ & n/a & n/a & n/a & 1 & +210.27 & n/a & n/a & n/a \\
H$_2$ & P & 4 & +9.56 & 21.59 & 0.2534 & 0.2534 & 3 & +23.67 & 22.59 & 0.0459 & 0.1098 \\
LiH & P & 4 & $-9.73$ & 4.80 & 0.0076 & 0.0302 & 4 & +3.52 & 2.78 & 0.0275 & 0.1098 \\
HF & P & 4 & +7.71 & 10.10 & 0.0932 & 0.1937 & 4 & +9.02 & 7.36 & 0.0299 & 0.1098 \\
O$_2$ & P & 5 & $-108.75$ & 119.28 & 0.0646 & 0.1937 & 5 & +99.50 & 100.99 & 0.0521 & 0.1098 \\
BeH$_2$ & P & 1 & +19.56 & n/a & n/a & n/a & 1 & +62.17 & n/a & n/a & n/a \\
H$_2$O & P & 1 & +78.13 & n/a & n/a & n/a & 1 & +45.88 & n/a & n/a & n/a \\
\bottomrule
\end{tabular}}
\par\medskip
\textbf{(b) Cross-backend Welch comparisons}\par\smallskip
\begin{tabular}{lrrrrrr}
\toprule
Molecule & $n_A$ & $n_P$ & Mean A (Ha) & Mean P (Ha) & Welch $p$ & Holm $p$ \\
\midrule
H$_2$ & 3 & 4 & -1.85302 & -1.85394 & 0.92470 & 0.92470 \\
LiH & 4 & 4 & -1.04981 & -1.03656 & 0.00019 & 0.00077 \\
HF & 3 & 4 & -1.92576 & -1.92060 & 0.16316 & 0.32632 \\
O$_2$ & 2 & 5 & -4.89627 & -4.97033 & 0.10375 & 0.31125 \\
\bottomrule
\end{tabular}
\caption{Panel (a) reports final-point remeasurement shifts and drift-differenced contrasts in mHa. Positive values denote higher energy on later remeasurement. Source-record means are independent units; Holm correction is applied separately to the four core tests on each backend. The Pittsburgh H$_2$ selection analysis uses three records because the fourth has identical source best and final energies but distinct vectors ($\lVert\Delta\theta\rVert_2=1.0\times10^{-3}$) and is reported as a zero-source-gap diagnostic. Across the 17 Pittsburgh best-point source vectors, the pooled mean paired shift is $+6.80$ mHa (two-sided paired $t_{16}=0.71$, $p=0.487$). Panel (b) gives absolute optimizer-selected mean energies in hartree for Aachen (A) and Pittsburgh (P); Welch tests use the same energies and only LiH remains resolved after Holm adjustment of the four-molecule family.}
\label{tab:theta-final-control}\label{tab:drift-differenced-selection}\label{tab:welch-backends}
\end{table}

\FloatBarrier
\begin{table*}[!ht]
\centering
\scriptsize
\setlength{\tabcolsep}{3.5pt}
\resizebox{\linewidth}{!}{%
\begin{tabular}{llrrrrrrrrr}
\toprule
Backend & Molecule & $n_d$ & Lag (d) & Mean $s_b^{(2)}$ & $\Delta E_{b,d}^{(2-1)}$ & $\Delta E_{f,d}^{(2-1)}$ & $d_{\rm sel}^{(1)}$ & $d_{\rm sel}^{(2)}$ & $\Delta d_{\rm sel}$ & Resolution \\
\midrule
Aachen & H$_2$ & 3 & +1.7 & +7.17 & +3.31 & +2.02 & +13.52 & +14.80 & +1.28 & +2.85 \\
Aachen & LiH & 4 & +1.6 & +1.59 & +0.49 & +1.11 & +3.79 & +3.17 & $-0.63$ & +0.60 \\
Aachen & HF & 3 & +1.6 & +4.93 & +2.20 & +2.83 & +6.44 & +5.81 & $-0.63$ & +1.48 \\
Aachen & O$_2$ & 2 & +1.5 & +13.97 & $-21.72$ & $-19.63$ & +25.13 & +23.03 & $-2.09$ & +9.06 \\
Pittsburgh & H$_2$ & 3 & +8.8 & +10.75 & $-8.78$ & $-10.93$ & +23.67 & +25.81 & +2.15 & +4.15 \\
Pittsburgh & LiH & 4 & +8.8 & +1.91 & +1.11 & +1.32 & +3.52 & +3.32 & $-0.20$ & +0.64 \\
Pittsburgh & HF & 4 & +8.8 & +5.47 & $-3.04$ & $-4.57$ & +9.02 & +10.54 & +1.52 & +1.96 \\
Pittsburgh & O$_2$ & 5 & +9.0 & +20.64 & $-29.08$ & $-31.29$ & +99.50 & +101.72 & +2.21 & +8.01 \\
Pittsburgh & H$_2$ (diagnostic) & n/a & +8.8 & +10.54 & n/a & n/a & +10.41 & +3.05 & $-7.36$ & +6.86 \\
\bottomrule
\end{tabular}}
\caption{Repeated resilience-0 measurements of identical saved parameter vectors in two later calibration epochs. Lag is the mean interval between fixed-parameter campaigns. Mean $s_b^{(2)}$ is the cell mean of per-record epoch-2 best-vector sample SDs. For $d_{\rm sel}/b$, the aggregate epoch-2 mean-of-record ratio 1.295, ratio of means 1.687, and old-to-new correlation 0.994 use archived per-record epoch-2 scatter rather than the cell means printed here. The per-record values reproducing these statistics and the variance ratio 0.38 are deposited in \texttt{epoch\_remeasurement\_comparison.csv}. The second-minus-first energy changes $\Delta E_{b,d}^{(2-1)}$ and $\Delta E_{f,d}^{(2-1)}$ are restricted to the predefined $d_{\rm sel}$ source set, so $\Delta d_{\rm sel}=\Delta E_{b,d}^{(2-1)}-\Delta E_{f,d}^{(2-1)}$ closes arithmetically. Resolution is the propagated one-standard-deviation uncertainty from the four within-vector sample means. Entries are cell means in mHa; repeated evaluations are not independent source runs. The Pittsburgh H$_2$ zero-source-drift record is shown separately as a zero-source-gap diagnostic and excluded from cell summaries. The successful Pittsburgh O$_2$ retry was split across two Runtime jobs after the original timed-out record, which remains preserved as a documented failure.}
\label{tab:epoch-remeasurement}
\end{table*}

\FloatBarrier

\begin{figure}[H]
    \centering
    \includegraphics[width=\linewidth]{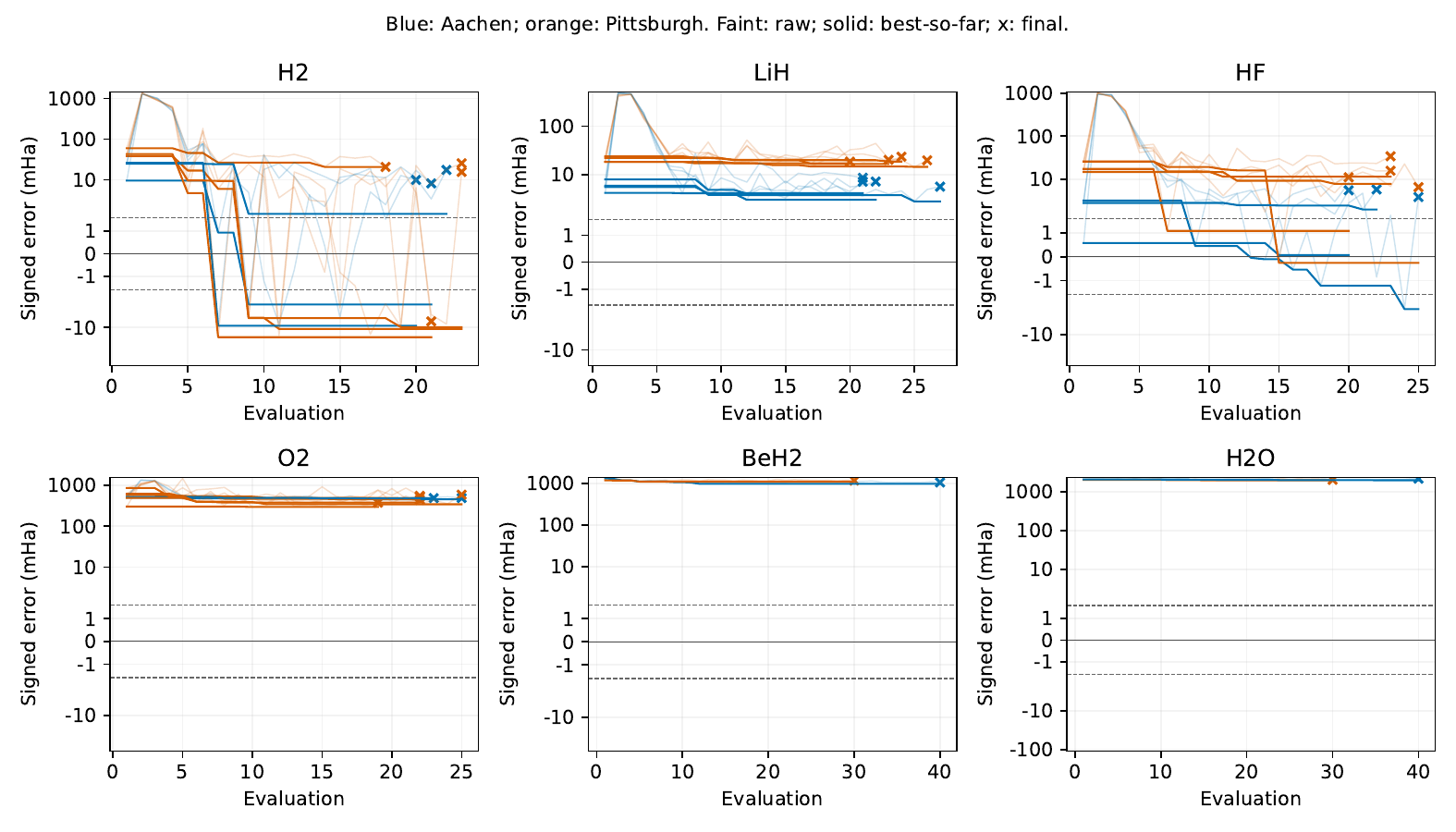}
    \caption{Hardware optimization traces as signed error relative to active-space CASCI. Faint lines are single noisy objective evaluations, solid lines are best-so-far samples and crosses identify final evaluations; no per-shot uncertainty is available. Each panel contains all successful source runs for one molecule. Dashed lines mark the 1 kcal mol$^{-1}$ band where visible.}
    \label{fig:optimizer-traces}
\end{figure}

\begin{figure}[H]
    \centering
    \includegraphics[width=0.82\linewidth]{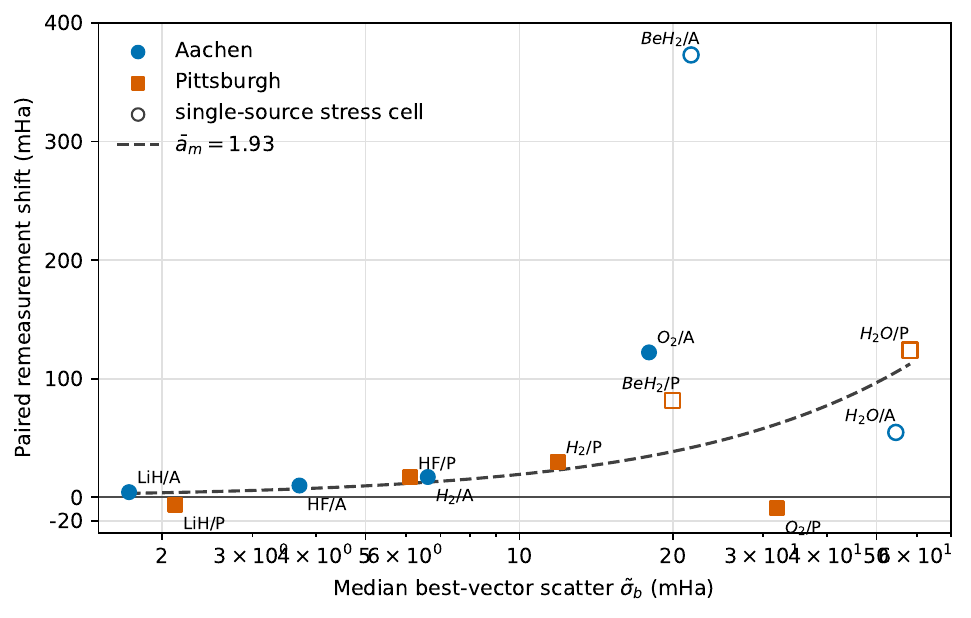}
    \caption{Cell-mean $\theta_\mathrm{best}$ remeasurement shift against the median within-vector best-point SD $\tilde\sigma_b$. The dashed line uses the median Gaussian order-statistic coefficient $\bar a_m=1.93$ across the 33 source records. No uncertainty bars are plotted. The all-cell association is descriptive; excluding Pittsburgh O$_2$ is post hoc.}
    \label{fig:selection-noise}
\end{figure}

\begin{table}[H]
\centering
\scriptsize
\textbf{(a) Clustered endpoint intervals}\par\smallskip
\resizebox{\linewidth}{!}{%
\begin{tabular}{llrrrrrrr}
\toprule
Set & Endpoint & Estimate & Cell-$t$ reference & Wild-cluster CI & Robust CI & IID infl. SE & Cluster SE & Containment margin\\
\midrule
core & $f_{\rm nr}$ (non-retained fraction) & 0.989 & $[0.865, 1.113]$ & $[0.906, 1.072]$ & $[0.891, 1.087]$ & 0.039 & 0.046 & $\pm$9.4\% \\
core & $\Delta_{fix}$ (mHa) & -0.919 & $[-3.30, 1.46]$ & $[-2.33, 0.49]$ & $[-2.50, 0.67]$ & 0.737 & 0.739 & n/a \\
core & $d_{sel}/b$ & 1.014 & $[0.782, 1.246]$ & $[0.853, 1.175]$ & $[0.836, 1.192]$ & 0.082 & 0.083 & n/a \\
extended & $f_{\rm nr}$ (non-retained fraction) & 0.972 & $[0.887, 1.057]$ & $[0.917, 1.027]$ & $[0.908, 1.036]$ & 0.026 & 0.030 & $\pm$8.3\% \\
extended & $\Delta_{fix}$ (mHa) & -0.887 & $[-5.65, 3.88]$ & $[-1.82, 0.05]$ & $[-1.93, 0.15]$ & 0.490 & 0.491 & n/a \\
extended & $d_{sel}/b$ & 1.111 & $[0.844, 1.378]$ & $[1.005, 1.218]$ & $[0.994, 1.228]$ & 0.054 & 0.055 & n/a \\
\bottomrule
\end{tabular}}
\par\smallskip
\textbf{(b) Precision-weight sensitivity}\par\smallskip
\begin{tabular}{llrrr}
\toprule
Backend & Molecule & $f_{\rm nr}$ & Record-ratio SE & Normalized weight \\
\midrule
Aachen & H2 & 0.883 & 0.010 & 0.853 \\
Pittsburgh & H2 & 0.982 & 0.134 & 0.005 \\
Aachen & LiH & 1.023 & 0.073 & 0.018 \\
Pittsburgh & LiH & 0.945 & 0.082 & 0.014 \\
Aachen & HF & 1.241 & 0.107 & 0.008 \\
Pittsburgh & HF & 0.763 & 0.073 & 0.017 \\
Aachen & O2 & 1.136 & 0.551 & 0.000 \\
Pittsburgh & O2 & 0.939 & 0.033 & 0.084 \\
\bottomrule
\end{tabular}
\caption{Panel (a): the cell-$t$ reference includes between-cell heterogeneity. The enumerated session-level wild-cluster (Rademacher) interval conditions on the benchmark cells and sign-flips cell-balanced record influence functions over complete sessions; no assignment drops a cell. Coverage is approximate; extended intervals condition on the four singleton stress observations, whose residual influences are zero. Against the comparable IID record-influence SE, clustering widens $f_{\rm nr}$ and $d_{\rm sel}/b$ uncertainty. Robust intervals use the session-sandwich SE with a $t_{G-1}$ critical value. The final column is the smallest symmetric margin around unity containing the reported 95\% $f_{\rm nr}$ interval. Panel (b): the sensitivity weights are inverse variances of record-level $f_{\rm nr}$ within each cell, normalized to sum to one; they are not weights based on $d_{\rm sel}$ precision.}
\label{tab:cluster-inference}
\end{table}

\FloatBarrier
\begin{table}[H]
\centering
\small
\begin{tabular}{lrrr}
\toprule
Optimizer & Dimension & Bias range / $\sigma$ & $-\mathrm{bias}/(a_m\sigma)$ range \\
\midrule
COBYLA & 1 & -2.35 to -1.97 & 0.96 to 1.05 \\
COBYLA & 2 & -2.48 to -1.86 & 0.97 to 1.04 \\
COBYLA & 4 & -2.54 to -1.74 & 0.93 to 1.02 \\
COBYLA & 8 & -2.73 to -1.69 & 0.90 to 1.04 \\
COBYLA & 16 & -3.04 to -1.68 & 0.90 to 1.06 \\
SPSA & 1 & -2.82 to -1.80 & 0.90 to 1.03 \\
SPSA & 2 & -2.94 to -1.73 & 0.87 to 0.97 \\
SPSA & 4 & -2.34 to -1.67 & 0.77 to 0.91 \\
SPSA & 8 & -1.97 to -1.44 & 0.58 to 0.84 \\
SPSA & 16 & -1.39 to -1.01 & 0.40 to 0.75 \\
\bottomrule
\end{tabular}
\caption{Offline quadratic-plus-Gaussian experiment using actual COBYLA and SPSA optimization. Each of the five budgets (20, 50, 100, 200, 500) has 24 independent fixed-seed replicates per optimizer and dimension. The objective is $E_{\rm true}(\boldsymbol\theta)=\|\boldsymbol\theta\|_2^2/2$ with independent Gaussian noise of SD $\sigma=1$ per evaluation. Bias is $E_{\rm best,observed}-E_{\rm true}(\boldsymbol\theta_{\rm best})$; $a_m$ uses the actual evaluation count. Ranges across budgets compress the full 50-row CSV. This is a generic optimizer selection experiment, not a device model or validation of the archived hardware contrast.}
\label{tab:synthetic-orderstat}
\end{table}

\FloatBarrier
\begin{table}[H]
\centering
\scriptsize
\begin{tabular}{lllrrr}
\toprule
Backend & Molecule & Source run & Best (Ha) & CASCI (Ha) & Below CASCI (mHa)\\
\midrule
pittsburgh & H2 & 2026-08-02-18.02.42 & $-1.867025010$ & $-1.849662051$ & 17.36 \\
pittsburgh & H2 & 2026-08-01-17.54.22 & $-1.860528194$ & $-1.849662051$ & 10.87 \\
pittsburgh & H2 & 2026.07.30-22.13.56 & $-1.859459764$ & $-1.849662051$ & 9.80 \\
aachen & H2 & 2026-08-01-18.57.40 & $-1.858766582$ & $-1.849662051$ & 9.10 \\
aachen & H2 & 2026-08-01-16.36.57 & $-1.852392301$ & $-1.849662051$ & 2.73 \\
aachen & HF & 2026-08-01-18.57.40 & $-1.928160708$ & $-1.925584360$ & 2.58 \\
pittsburgh & HF & 2026-08-01-12.09.37 & $-1.925841858$ & $-1.925584360$ & 0.26 \\
\bottomrule
\end{tabular}
\caption{Successful run-level optimizer-selected energies below the exact active-space CASCI reference. The final column is the positive magnitude below CASCI; these values cannot be interpreted as variational upper bounds.}
\label{tab:sub-casci-audit}
\end{table}

\FloatBarrier
\begin{table}[!ht]
\centering
\scriptsize
\resizebox{\linewidth}{!}{%
\begin{tabular}{llllllll}
\toprule
Session & Run ID & Backend & Molecule & $||\Delta\theta||_2$ & $\Delta E_{ideal}$ & $\Delta E_{fix}$ & $\Delta E_{fix}-\Delta E_{ideal}$\\
\midrule
A1 & 2026.07.30-21.17.21 & aachen & LiH & $1.0\times10^{-4}$ & $-0.0004$ & $-0.018$ & $-0.017$ \\
A2 & 2026-08-01-09.02.35 & aachen & H2 & $1.0\times10^{-4}$ & $-0.0204$ & $-2.078$ & $-2.058$ \\
A2 & 2026-08-01-09.02.35 & aachen & HF & $1.0\times10^{-4}$ & +0.0083 & +1.817 & +1.809 \\
A2 & 2026-08-01-09.02.35 & aachen & LiH & $1.0\times10^{-4}$ & +0.0021 & +0.543 & +0.541 \\
A3 & 2026-08-01-12.11.09 & aachen & BeH2 & $1.0\times10^{-2}$ & $-0.4843$ & $-7.796$ & $-7.312$ \\
A3 & 2026-08-01-12.11.09 & aachen & H2O & $1.0\times10^{-2}$ & +1.4152 & +19.679 & +18.264 \\
A3 & 2026-08-01-12.11.09 & aachen & O2 & $1.0\times10^{-4}$ & +0.0067 & +9.483 & +9.476 \\
A4 & 2026-08-01-16.36.57 & aachen & H2 & $1.0\times10^{-4}$ & +0.0204 & $-1.416$ & $-1.437$ \\
A4 & 2026-08-01-16.36.57 & aachen & HF & $1.0\times10^{-4}$ & +0.0060 & +1.007 & +1.001 \\
A4 & 2026-08-01-16.36.57 & aachen & LiH & $1.0\times10^{-4}$ & +0.0109 & $-0.669$ & $-0.680$ \\
A4 & 2026-08-01-16.36.57 & aachen & O2 & $1.0\times10^{-4}$ & +0.0367 & $-3.479$ & $-3.516$ \\
A5 & 2026-08-01-18.57.40 & aachen & H2 & $1.0\times10^{-4}$ & +0.0043 & $-1.890$ & $-1.895$ \\
A5 & 2026-08-01-18.57.40 & aachen & HF & $1.0\times10^{-4}$ & +0.0052 & +0.930 & +0.925 \\
A5 & 2026-08-01-18.57.40 & aachen & LiH & $1.0\times10^{-4}$ & +0.0021 & +0.489 & +0.487 \\
P1 & 2026.07.30-22.13.56 & pittsburgh & H2 & $1.0\times10^{-4}$ & $-0.0277$ & $-6.403$ & $-6.375$ \\
P2 & 2026.07.31-03.08.12 & pittsburgh & LiH & $1.0\times10^{-4}$ & +0.0062 & +0.351 & +0.345 \\
P3 & 2026.07.31-18.20.56 & pittsburgh & HF & $1.0\times10^{-4}$ & +0.0329 & $-6.968$ & $-7.001$ \\
P4 & 2026-08-01-12.09.37 & pittsburgh & H2 & $1.0\times10^{-3}$ & $-0.1117$ & +10.405 & +10.517 \\
P4 & 2026-08-01-12.09.37 & pittsburgh & HF & $1.0\times10^{-4}$ & $-0.0108$ & $-1.605$ & $-1.594$ \\
P4 & 2026-08-01-12.09.37 & pittsburgh & LiH & $1.0\times10^{-4}$ & +0.0094 & $-0.416$ & $-0.426$ \\
P5 & 2026-08-01-17.54.22 & pittsburgh & H2 & $1.0\times10^{-4}$ & $-0.0118$ & +2.231 & +2.243 \\
P5 & 2026-08-01-17.54.22 & pittsburgh & HF & $1.0\times10^{-4}$ & $-0.0009$ & +0.115 & +0.116 \\
P5 & 2026-08-01-17.54.22 & pittsburgh & LiH & $1.0\times10^{-4}$ & $-0.0006$ & $-0.388$ & $-0.388$ \\
P6 & 2026-08-01-18.56.23 & pittsburgh & O2 & $1.0\times10^{-4}$ & +0.0540 & +0.843 & +0.789 \\
P7 & 2026-08-02-07.22.35 & pittsburgh & O2 & $1.0\times10^{-4}$ & +0.0766 & $-9.853$ & $-9.929$ \\
P8 & 2026-08-02-16.04.58 & pittsburgh & O2 & $1.0\times10^{-4}$ & +0.0506 & $-3.393$ & $-3.443$ \\
P9 & 2026-08-02-18.02.42 & pittsburgh & H2 & $1.0\times10^{-4}$ & $-0.0118$ & +2.894 & +2.905 \\
P9 & 2026-08-02-18.02.42 & pittsburgh & HF & $1.0\times10^{-4}$ & $-0.0200$ & $-2.737$ & $-2.717$ \\
P9 & 2026-08-02-18.02.42 & pittsburgh & LiH & $1.0\times10^{-4}$ & $-0.0027$ & $-0.374$ & $-0.371$ \\
P9 & 2026-08-02-18.02.42 & pittsburgh & O2 & $1.0\times10^{-4}$ & $-0.1195$ & $-12.936$ & $-12.817$ \\
P10 & 2026-08-15-22.13.13 & pittsburgh & O2 & $1.0\times10^{-4}$ & $-0.0096$ & $-7.003$ & $-6.994$ \\
P11 & 2026-09-01-17.16.37 & pittsburgh & H2O & $5.0\times10^{-3}$ & +0.4625 & $-8.477$ & $-8.939$ \\
P12 & 2026-09-02-06.17.58 & pittsburgh & BeH2 & $1.0\times10^{-2}$ & +0.6753 & $-6.700$ & $-7.376$ \\
\bottomrule
\end{tabular}}
\caption{Record-level noiseless and hardware fixed-parameter gaps, all in mHa except the dimensionless parameter-vector norm. $\Delta E_{ideal}=E_{ideal}(\theta_{best})-E_{ideal}(\theta_{final})$; $\Delta E_{fix}$ uses the corresponding later-epoch hardware means. Pittsburgh/H$_2$ run 2026-08-01-12.09.37 is the predefined zero-source-drift zero-source-gap diagnostic and is excluded from the primary selection set. The median absolute ideal gap is \idealGapMedian{} mHa and the range is \idealGapMin{}--\idealGapMax{} mHa. The maximum is the Aachen H$_2$O record 2026-08-01-12.11.09; that cell also has the largest $d_{sel}$, 210.27 mHa.}
\label{tab:ideal-gap-control}
\end{table}

\FloatBarrier
\begin{table}[H]
\centering
\small
\begin{tabular}{rrrr}
\toprule
Orbital & Energy (Ha) & Occupation & Irrep \\
\midrule
6 & $-0.575817682$ & 2.0 & E1ux \\
7 & $-0.073470272$ & 1.0 & E1gx \\
8 & $-0.073470272$ & 1.0 & E1gy \\
9 & 0.720396859 & 0.0 & A1u \\
\bottomrule
\end{tabular}
\caption{PySCF ROHF orbital audit for the O$_2$/STO-3G active window. Zero-indexed orbitals 7 and 8 are degenerate to $10^{-8}$ Ha and retain both components of the $\pi_g^*$ pair. The compact window preserves this antibonding degeneracy.}
\label{tab:o2-symmetry}
\end{table}

\FloatBarrier
\begin{table}[H]
\centering
\small
\begin{tabular}{lr}
\toprule
Parameter & Value \\
\midrule
Backend / molecule & Phoenix / LiH \\
Processor / active space & Nighthawk r2 / 2e2o \\
Independent records / realized $m$ & 5 / 19--22 \\
Mean $a_m$ / source drift (mHa) & 1.889 / 6.21 \\
Non-retained fraction, ratio of means & 1.060 \\
Non-retained fraction, mean of record ratios & 1.024 [0.871, 1.177] \\
$d_{sel}$ (mHa) / $d_{sel}/b$ mean of ratios & 6.59 / 2.04 \\
$d_{sel}/b$ ratio of means / best error (mHa) & 2.11 / +11.50 \\
Shots / resilience / mapper & 1024 / 0 / PT \\
Qubits / depth / CZ / basis & 2 / 43 / 4 / cz,id,rz,sx,x \\
\bottomrule
\end{tabular}
\caption{Dated, post hoc Phoenix LiH cross-family observation. Energies and shifts are in mHa. RoM denotes ratio of means; MoR denotes mean of record ratios. The record-level $t$ interval uses five job-mode records and $df=4$; no interval is assigned to RoM with five clusters. The purported timing of the $\geq1.5$ criterion is not independently verified, so it is not used as a confirmatory rule. Exact two-sided sign enumeration has minimum $p=0.0625$. The cell is standalone. The fixed-job $d_{sel}/b$ MoR and RoM are 2.04 and 2.11; source-plateau scatter gives 1.25 and 1.17, respectively.}
\label{tab:phoenix-control}
\end{table}

\FloatBarrier
\begin{table}[H]
\centering
\small
\begin{tabular}{lrl}
\toprule
Disposition or control & Count & Endpoint treatment \\
\midrule
Retained Heron source records (A/P) & 33 (14/19) & 32 contrast records \\
Failed source records & 2 & Excluded \\
Running source records & 4 & Excluded \\
Missing-status source record & 1 & Excluded \\
Exact duplicate copies & 2 & Excluded copies only \\
Failed-recovered source marker & 1 & Excluded marker \\
Post hoc Phoenix source records & 5 & Standalone cohort \\
Superseded Phoenix completed jobs & 2 & Payload unavailable \\
Timed-out O$_2$ fixed job & 1 & Failed; retry retained \\
\bottomrule
\end{tabular}
\caption{Source-record dispositions come from the saved integrity ledger; fixed-job and Phoenix job rows are separate accounting levels and must not be summed with source records. The retained Pittsburgh H$_2$ zero-source-gap record enters raw-shift summaries but not the primary contrast ratio. No running or missing-status source enters a quantitative endpoint.}
\label{tab:inclusion-flow}
\end{table}

\FloatBarrier
\clearpage
\subsection*{Software provenance}
Local package stamps and remote job labels are distinct (Supplementary Table~\ref{tab:software-provenance}): shallow-source remote metrics report Qiskit 2.3.0, while their local stamp is 1.4.3. The records do not identify when or why the local versions changed. Version was not assigned as an experimental condition, and its hardware effect cannot be separated from acquisition epoch and task. The offline analysis environment is reported separately in Methods.
\begin{table}[H]
\centering
\small
\begin{tabular}{lrlllll}
\toprule
Records & $n$ & Qiskit & Runtime & Aer & Algorithms & Nature \\
\midrule
Source runs & 31 & 1.4.3 & 0.40.1 & 0.17.1 & 0.3.1 & 0.7.2 \\
Source runs & 7 & 2.5.2 & 0.49.0 & 0.17.2 & 0.4.0 & 0.7.2 \\
Fixed acquisitions & 143 & 2.5.2 & 0.49.0 & 0.17.2 & 0.4.0 & 0.7.2 \\
\bottomrule
\end{tabular}
\caption{Local package stamps copied from archived run/acquisition metadata. Counts are source calculations or fixed acquisitions, not independent sampling units. Per-record links are supplied in \texttt{software\_provenance.json} with the dataset.}\label{tab:software-provenance}
\end{table}

\FloatBarrier
The paired local control reconstructs the same Hamiltonians and saved parameter vectors under the two SDK sets, holding Qiskit Nature and the classical dependencies fixed. Logical states, mapped Hamiltonians, ideal energies and compact noisy expectations agree numerically (Supplementary Table~\ref{tab:software-compatibility}). This is a present-day sensitivity calculation, not a replay of complete historical environments: unrecorded active-orbital indices use the same automatic selection in both controls and remain null in the archive. Synthetic-target gate counts and depths differ for the deep stress circuits despite agreement of ideal expectations. The legacy client EstimatorOptions declarations also agree, but this does not establish identical resolved remote options. These checks support compatibility of the reconstructed local calculations, not equivalence of historical hardware execution or additional independent samples.
\begin{table}[H]
\centering
\small
\begin{tabular}{lrr}
\toprule
Local comparison & Points/models & Maximum difference \\
\midrule
Pauli coefficients (Ha) & 6 & 0 \\
Logical energies (Ha) & 854 & 1.07e-14 \\
Logical state infidelity & 854 & 3.33e-16 \\
Synthetic-target energies (Ha) & 854 & 3.11e-14 \\
Compact noisy energies (Ha) & 76 & 2.66e-15 \\
\bottomrule
\end{tabular}
\caption{Deterministic paired local comparison of the SDK versions in Supplementary Table \ref{tab:software-provenance}, using the same saved vectors and common classical dependencies. Counts indicate functional coverage, not additional hardware samples. Noise uses the archived compact depolarizing models. The five-qubit line target is synthetic, with identity initial layout, seed 1234 and level-0 transpilation; it does not reproduce historical routing or calibration. Native gate counts and depths match for H$_2$, LiH, HF and O$_2$ but differ for BeH$_2$ and H$_2$O. Remote execution and resolved server options are not tested.}\label{tab:software-compatibility}
\end{table}

\FloatBarrier
\end{document}